\documentclass[%
 aip,
 amsmath,amssymb,
reprint,
]{revtex4-1}

\usepackage{graphicx}
\usepackage{dcolumn}
\usepackage{bm}

\usepackage[dvipsnames]{xcolor}
\usepackage{hyperref}
\hypersetup{
    colorlinks  =   true,
    citecolor   =   BrickRed,
    linkcolor   =   RoyalBlue,
    urlcolor    =   Black 
}

\usepackage[utf8]{inputenc}
\usepackage[T1]{fontenc}
\usepackage{mathptmx}
\usepackage{etoolbox}
\usepackage{amsmath}
\usepackage{amssymb}
\usepackage{bbold}

\usepackage{float}
\usepackage{graphicx}
\usepackage{dcolumn}
\usepackage{bm}
\usepackage{bbm}
\usepackage{mathrsfs}

\usepackage{tikz}
\usetikzlibrary{matrix}
\usepackage{tikz-cd}

\newcommand{\la}{\langle}
\newcommand{\ra}{\rangle}
\newcommand{\tp}{\mathsf{T}}
\newcommand{\ex}{\text{ex}}

\DeclareMathOperator{\per}{per}
\DeclareMathOperator{\haf}{haf}
\DeclareMathOperator{\tr}{tr}

\makeatletter
\def\@email#1#2{%
 \endgroup
 \patchcmd{\titleblock@produce}
  {\frontmatter@RRAPformat}
  {\frontmatter@RRAPformat{\produce@RRAP{*#1\href{mailto:#2}{#2}}}\frontmatter@RRAPformat}
  {}{}
}%
\makeatother
\begin{document}


\title[]{Towards the simplest model of quantum supremacy:\\ Atomic boson sampling in a box trap}

\affiliation{Department of Physics and Astronomy and Institute for Quantum Science and Engineering, Texas A\&M University, College Station, TX 77843, USA}
\author{V. V. Kocharovsky} 
\email{vkochar@physics.tamu.edu}
\author{Vl. V. Kocharovsky}
\author{W. D. Shannon}
\author{S. V. Tarasov}

\date{\today}

\begin{abstract}
We describe boson sampling of interacting atoms from the noncondensed fraction of Bose-Einstein-condensed (BEC) gas confined in a box trap as a new platform for studying computational $\sharp$P-hardness and quantum supremacy of many-body systems. 
We calculate the characteristic function and statistics of atom numbers via newly found hafnian master theorem. 
Using Bloch-Messiah reduction, we find that interatomic interactions give rise to two equally important entities -- eigen-squeeze modes and eigen-energy quasiparticles -- whose interplay with sampling atom states determines behavior of the BEC gas. 
We infer that two necessary ingredients of $\sharp$P-hardness, squeezing and interference, are self-generated in the gas and, contrary to Gaussian boson sampling in linear interferometers, external sources of squeezed bosons are not required.
\end{abstract}

\maketitle

\author{V. V. Kocharovsky$^a^)$, Vl. V. Kocharovsky, W. D. Shannon, and S. V. Tarasov\\
$^a^)$ \textbf{Electronic mail:} vkochar@physics.tamu.edu}

\maketitle

\section{Introduction: The simplest quantum many-body model showing $\sharp$P-hard complexity}

Analysis of various quantum many-body systems capable of simulating $\sharp$P-hard computational problems is one of the main topics of modern research in quantum physics and computing (see, for example, \cite{Aaronson2011,Aaronson2013,Harrow2017,Brod2019,Zhong2020,Boixo2018,Arute2019,Dalzell2020} and references therein). 
Recently, an atomic boson sampling of excited atom occupations in an equilibrium gas with a Bose-Einstein condensate (BEC) has been suggested as a process that could be $\sharp$P-hard for classical computing \cite{PRA2022}. 
An example of a multi-qubit BEC trap \cite{Entropy2022} shows that it could serve as a rich and, at the same time, convenient platform for studies of various phenomena associated with atomic boson sampling and quantum supremacy. 

The present paper aims at the simplest possible model of the BEC trap that would allow one to greatly simplify the general theory outlined in \cite{PRA2022,Entropy2022} and explicitly disclose the mechanism behind the $\sharp$P-hard computational complexity of quantum many-body systems. 
As a result, many general formulas and the entire theory, based on the method of characteristic function and hafnian master theorem, acquire an explicit, transparent form clearly revealing the origin of the $\sharp$P-hardness of computing the joint probability distribution of the excited-atom occupations. 
This $\sharp$P-hardness is the ultimate reason for a potential quantum supremacy of the atomic boson sampling over classical-computing-based implementations of generation of such random numbers. 
Remarkably, as is shown in section VIII, the $\sharp$P-hard complexity of quantum systems is equivalent to an intuitively obvious complexity of computing the multivariate integral for Fourier series coefficients of a sign-indefinite strongly-oscillating function.

Due to the interatomic interaction, nonzero mass of atoms, equilibrium, and an absence of external sources of bosons the proposed atomic boson sampling has substantially different physics compared to photonic boson sampling in a linear interferometer widely studied in the last decade, both theoretically 
\cite{Aaronson2013, Harrow2017, Brod2019, Hamilton2017, HamiltonPRA2019, Quesada2022, Lim2022, LundPRL2014, Shi2021, Shchesnovich2019, Chin2018, Quesada2018, Huh2019, Huh2020,Villalonga2021,BentivegnaBayesianTest2015,Renema2018,Renema2020,Popova2021,Qi2020,Lund2015} 
and experimentally 
\cite{Zhong2020, Bentivegna2015, Wang2017,Loredo2017, Zhong2019,Wang2019,PanPRL2021,Madsen2022}.
The interatomic interaction is especially important. 
It greatly complicates and changes behavior of the quantum system of many bosons leading, in particular, to the fundamental phenomenon of two-mode squeezing of excited atom states predicted in \cite{PRA2000} and strongly pronounced in the statistics of the total noncondensate occupation \cite{PRA2020}. 
Nevertheless, simplicity of the BEC-in-the-box model allows us to disclose analytically the quantum-statistical physics of sampling the interacting Bose atoms. 
Our results show that the joint probability distribution of atom numbers in atomic boson sampling is expressed via the matrix hafnian in a way similar to that of the Gaussian boson sampling which utilizes external sources of squeezed photons. 
Thus, the same $\sharp$P-hardness of computing the hafnian is involved. 
It allows one to transfer a large part of the previously developed analysis of the computational complexity and quantum supremacy from photonic to atomic boson sampling. 

In addition to fundamentals of quantum simulations, physics of occupation fluctuations in the excited atom states of the BEC gas is very important for various applications and other BEC-gas setups, for instance, for studying trap cells \cite{Castin,Pit2011}, BEC collapse \cite{Calzetta2003}, squeezed states \cite{DrummondPRA2019}, matter-wave interferometers \cite{Shin2004}, including Ramsey \cite{DrummondPRA2019,Drummond2011} and Mach-Zehnder on-chip \cite{Chip1000atoms} ones, etc.  

The content of the paper is as follows. 
In section~II we introduce the main quantities and notations as well as summarize some known facts relevant to the many-body BEC system of atoms in a box trap. 
In sections~III and IV we present the general solution for sampling probabilities obtained by means of the hafnian master theorem and reveal two ingredients of quantum supremacy, squeezing and interference, via the Bloch-Messiah reduction of the Bogoliubov transformation. 
In sections V through VII we illustrate increasing apparent complexity of sampling probability patterns with changing the observational basis of excited atom states from the eigen-squeeze modes to more and more involved unitary mixtures of them. 
Section~VIII contains discussion of the $\sharp$P-hardness of computing the joint atom-number probabilities in a general case of arbitrary sampling states. 
The focus is on the origin of $\sharp$P-hard complexity of atomic boson sampling due to interference and squeezing of the sampled atom states via their interplay with the eigen-squeeze modes and eigen-energy quasiparticles. 
Concluding remarks related to a possible experimental demonstration of atomic boson sampling and various manifestations of quantum supremacy constitute section~IX. 

\section{Quantum statistical physics of joint fluctuations of the excited atom numbers}

Consider a box trap of volume $V=L^3$  with periodic boundary conditions. 
Excited atoms are described by a field operator 
\begin{equation} \label{field}
\hat\psi_\ex({\bf r}) = \sum_l \ \phi_l({\bf r}) \hat{b}_l \ ,
\end{equation}
where $\hat{b}_l$ denotes an operator annihilating an atom in the bare-atom excited state $\phi_l({\bf r})$ of the single-particle Hilbert space $\mathcal{H}$. 
The sum $\sum_l$ denotes a summation over a basis of excited atom states in a box trap, $\{\phi_l\}$, enumerated by an integer $l \neq 0$. 

A condensate wave function in the box trap corresponds to the zero integer $l=0$ and is uniform, $\phi_0 = V^{-1/2}$. We abide the Bogoliubov-Popov approximation \cite{Shi1998,Zagrebnov2001} and replace the condensate annihilation operator by a c-number, $\hat{b}_0 \approx \sqrt{N_0}$, where $N_0$ is a mean number of atoms in the condensate.

The Bogoliubov Hamiltonian in the basis $\{\phi_l\}$ is given, up to an insignificant additive c-valued constant, by a quadratic form in the creation and annihilation operators: 
\begin{equation} \label{HH}
\begin{split}
&\hat{H} = \frac{1}{2} \left(  \begin{matrix}  
                            \hat{{\bf b}}^\dagger \\ 
                            \hat{{\bf b}}
                        \end{matrix} \right)^\tp
                    \! H 
                    \left(  \begin{matrix}  
                            \hat{{\bf b}}^\dagger \\ 
                            \hat{{\bf b}}
                        \end{matrix} \right), \qquad H =\left[ \begin{matrix}
                                \tilde{\Delta}  &   \epsilon + \Delta\\
                                \epsilon + \Delta^*   &  \tilde{\Delta}^*     \end{matrix} \right], \\
&\epsilon = \Big(\langle \phi_l | \hat{\epsilon} | \phi_{l'} \rangle \Big), \quad \ \hat{\epsilon} = \hbar^2 \nabla^2/(2m) \ , \\
&\Delta = \Big( gN_0 \int \phi_l^* |\phi_0|^2 \phi_{l'} \, d^3{\bf r} \Big), \ \ \tilde{\Delta} = \Big( gN_0 \int \phi_l^* \phi_{l'}^* \, \phi_0^2 \, d^3{\bf r} \Big).
\end{split}
\end{equation}
It is written via a ($2\times 2$)-block matrix $H$ which is built of matrices $\epsilon, \ \Delta$, and $\tilde{\Delta}$ as its blocks, applied to a 2-block column vector $\left(  \begin{matrix} \hat{{\bf b}}^\dagger \\ \hat{{\bf b}} \end{matrix} \right)$, consisting of the column vector ${\bf \hat{b}}^\dagger$ on top of the column vector ${\bf \hat{b}}$, and multiplied from the left by a 2-block row vector $({\bf \hat{b}}^\dagger, {\bf \hat{b}}) = \left( \begin{matrix} \hat{{\bf b}}^\dagger \\ \hat{{\bf b}} \end{matrix} \right)^\tp$. The bold-faced operator ${\bf \hat{b}} = \{\hat{b}_l \}^\tp$ or ${\bf \hat{b}}^\dagger = \{\hat{b}_l^\dagger \}^\tp$ denotes a column vector of annihilation or creation operators in the $\{\phi_l\}$ basis; $\hat{\epsilon} = \hbar^2 \nabla^2/(2m)$ is the single-particle energy operator for a bare atom of mass $m$ in the box trap. The nabla symbol $\nabla$ stands for the 3D-vector differential operator. The interatomic interaction is determined by a constant $g = 4\pi \hbar^2 a_s/m$ via the s-wave scattering length $a_s$. 

The general notations for the field operator and Hamiltonian in Eqs.~(\ref{field}), (\ref{HH}) are convenient for an abstract general analysis presented below in sections III, IV, and VIII. For any particular choice of the basis of excited atom states $\{ \phi_l \}$ the field operator and Hamiltonian acquire more specific and transparent form. In a usual basis of plane traveling waves 
\begin{equation} \label{expikr}
\phi_{\bf k} = \frac{e^{i{\bf kr}}}{\sqrt{V}}, \qquad {\bf k} = \frac{2\pi {\bf j}}{L}, 
\quad {\bf j} = \{ j_x,j_y,j_z \} \in \mathbb{N}^3,
\end{equation}
with the wave vectors ${\bf k}$ enumerated by the integer 3-vector ${\bf j}$, we have 
\begin{equation} \label{field-k}
    \hat\psi_\ex({\bf r}) = \sum_{{\bf k}\neq 0} \frac{e^{i {\bf k r}}}{\sqrt{V}} \ \hat{a}_{\bf k},
\end{equation}
\begin{equation} \label{H-k}
\begin{split}
&\hat{H} = \frac{1}{2} \left(  \begin{matrix}  
                            \hat{{\bf a}}^\dagger \\ 
                            \hat{{\bf a}}
                        \end{matrix} \right)^\tp
                    \! \left[ \begin{matrix}
                                \tilde{\Delta}  &   \epsilon + \Delta\\
                                \epsilon + \Delta^*   &  \tilde{\Delta}^*     \end{matrix} \right]
                    \left(  \begin{matrix}  
                            \hat{{\bf a}}^\dagger \\ 
                            \hat{{\bf a}}
                        \end{matrix} \right), \qquad \epsilon_{{\bf k}} = \frac{\hbar^2 {\bf k}^2}{2m}, \\
&\epsilon_{{\bf k}{\bf k'}}=\epsilon_{{\bf k}} \delta_{{\bf k}{\bf k'}}, \quad
\Delta_{{\bf k},{\bf k'}} = gN_0 \delta_{{\bf k},{\bf k'}}, \quad
    \tilde{\Delta}_{{\bf k},{\bf k'}} = gN_0 \delta_{{\bf k},{\bf -k'}}.
\end{split}
\end{equation}
Hereinafter, we denote the creation and annihilation operators of bare atoms in the traveling-plane-wave states $\frac{e^{i{\bf kr}}}{\sqrt{V}}$ by symbols $\hat{\bf a}^\dagger = \{\hat{a}_{\bf k}^\dagger |\, {\bf k} \neq 0 \}^\tp$ and $\hat{\bf a} = \{ \hat{a}_{\bf k}|\, {\bf k} \neq 0 \}^\tp$, respectively, as opposed to the above general-case operators $\hat{{\bf b}}^\dagger$ and $\hat{{\bf b}}$. 

Equilibrium quantum many-body statistics of atoms in the BEC trap at a temperature $T$ is determined by the statistical operator $\hat{\rho} = e^{-\hat{H}/T} \big/  \, \tr \big\{ e^{-\hat{H}/T} \big\}$. The atomic boson sampling implies sampling in accord with the joint probability distribution $\rho\big(\{ n_l \}\big)$ of the occupation numbers $\{n_l \}$ of the excited atom states $\{\phi_l \}$ in Eq.~(\ref{field}) or some subset or groups of them preselected for measurenment by the appropriate detectors projecting atoms onto those states. This probability distribution is given by Fourier series coefficients,
\begin{equation} \label{PDF} 
\rho\big(\{ n_l \}\big) = \int_{-\pi}^{\pi}...\int_{-\pi}^{\pi} e^{-i \sum_l \tau_l n_l} \ \Theta \big(\{ \tau_l \}\big) \prod_l \frac{d\tau_l}{2\pi}, \ n_l = 0,1,\ldots,
\end{equation} 
of the characteristic function of the atom-number operators:
\begin{equation} \label{defCF} 
\Theta\big(\{ \tau_l \}\big) = \big\la e^{i\sum_l \tau_l\hat{n}_l} \big\ra 
\equiv \tr \big\{ e^{i\sum_l \tau_l\hat{n}_l} \hat{\rho} \big\}, \quad \hat{n}_l = \hat{b}_l^\dagger \hat{b}_l \ . 
\end{equation} 
The symbol $\tr\{\ldots\}$ stands for a trace of an operator or matrix. 

The characteristic function (\ref{defCF}) had been found analytically \cite{PRA2022} via a determinant function which can be easily computed in polynomial time for any finite-size matrix:
\begin{equation} \label{CFdet}
\Theta = \frac{1}{\sqrt{\det (\mathbbm{1} + G - ZG)}}; \quad
Z = \begin{bmatrix}
        \textrm{diag} \big(\{ z_l \}\big) & \mathbb{0} \\    
        \mathbb{0} & \textrm{diag} \big(\{ z_l \}\big)
    \end{bmatrix}.
\end{equation}
Here the symbol $\mathbb{0}$ stands for the matrix block with zero entries.

The characteristic function is determined by a covariance matrix 
\begin{equation} \label{G=x}
G = \bigg\langle \!\! : \!  \left( \begin{matrix} \hat{{\bf b}}^\dagger \\ \hat{{\bf b}} \end{matrix} \right) 
\bigotimes \ ({\bf \hat{b}}, {\bf \hat{b}}^\dagger) \! : \!\! \bigg\rangle
= \left[ \begin{matrix}
            \big(\la \hat{b}_l^\dagger \hat{b}_{l'} \ra\big) 
            &   
            \big(\la \hat{b}_l^\dagger \hat{b}_{l'}^\dagger \ra\big)
            \\[6pt]
            \big(\la \hat{b}_l \hat{b}_{l'} \ra\big) 
            &   
            \big(\la \hat{b}_l^\dagger \hat{b}_{l'} \ra\big)
        \end{matrix} \right]
\end{equation}
whose entries are given by the quantum-statistical average of the normally ordered tensor product of the 2-block column vector $\left( \begin{matrix} \hat{{\bf b}}^\dagger \\ \hat{{\bf b}} \end{matrix} \right)$ and 2-block row vector $({\bf \hat{b}}, {\bf \hat{b}}^\dagger)$ of the atom creation and annihilation operators, that is, all possible self- and inter-mode normal, $\la \hat{b}_l^\dagger \hat{b}_{l'} \ra$, and anomalous, $\la \hat{b}_l^\dagger \hat{b}_{l'}^\dagger \ra$, $\la \hat{b}_l \hat{b}_{l'} \ra$, correlators. 
Each variable $\tau_l$ of the characteristic function $\Theta(\{ \tau_l\})$ appears in the matrix of variables $Z$ in Eq.~(\ref{CFdet}) twice --- via the entry $z_l = e^{i\tau_l}$ in each of the two identical diagonal blocks $\textrm{diag} \{ z_l \}$; the symbol $\mathbbm{1}$ denotes the identity matrix. Note that in the literature on Gaussian states \cite{Weedbrook2012}, the covariance matrix is often defined with a half anti-commutator, $(\hat{b}_l^\dagger \hat{b}_{l'} + \hat{b}_{l'} \hat{b}_l^\dagger)/2$, replacing the normal product, $\hat{b}_l^\dagger \hat{b}_{l'}$, of the creation/annihilation operators, that adds a half identity matrix $\mathbbm{1}/2$ to our covariance matrix $G$.

The characteristic function in Eq.~(\ref{CFdet}) has the same form for an arbitrary restricted, marginal subset of the excited states or coarse-grained groups of them, if they are considered irrespective to the other excited states. In order to average over all irrelevant excited-state occupations one just need to nullify all irrelevant variables $z_{l'}$ and keep only those rows and columns in the covariance matrix $G$ which correspond to the chosen marginal subset of excited states. 
In order to combine some excited states into a coarse-grained group it is necessary to set equal all of the variables $\{ z_l \}$ within such a group.

Consider the quasiparticle creation and annihilation operators, which constitute the column vectors $\hat{\tilde{{\bf b}}}^\dagger = \{\hat{\tilde{b}}_l^\dagger \}^\tp$ and $\hat{\tilde{{\bf b}}} = \{ \hat{\tilde{b}}_l \}^\tp$ and diagonalize the Hamiltonian in Eq.~(\ref{HH}),
\begin{equation} \label{QPH}
\hat{H} = \sum_l E_l \hat{\tilde{b}}_l^\dagger \hat{\tilde{b}}_l. 
\end{equation}
The quasiparticle eigen energy is denoted as $E_l$ and is given in Eq.~(\ref{diagH-k}) or (\ref{eigen-squeeze energies}).  
The field operator of excited atoms in Eq.~(\ref{field}) also can be represented via the quasiparticle operators as a sum of both annihilation and creation quasiparticle operators: 
\begin{equation} \label{field:ba-qp}
\hat\psi_\ex({\bf r}) = \sum_l \Big(\, u_l ({\bf r}) \hat{\tilde{b}}_l + v_l^* ({\bf r}) \hat{\tilde{b}}_l^\dagger \,\Big) \ .
\end{equation}
The functions $u_l ({\bf r})$ and $v_l^* ({\bf r})$ constitute the two-component wave function of the $l$-th quasiparticle. 
Canonical Bose commutation relations for the quasiparticle operators, $\hat{\tilde{b}}_l \hat{\tilde{b}}_{l'}^\dagger - \hat{\tilde{b}}_l^\dagger \hat{\tilde{b}}_{l'} = \delta_{l,l'}$, where $\delta_{l,l'}$ is the Kronecker delta, imply the following normalization of the two components, $u_l$ and $v_l^*$, of each quasiparticle wave function
\begin{equation} \label{norm}
\int_V \left(|u_l|^2 - |v_l|^2 \right) \, d^3 {\bf r} = 1 .
\end{equation}

Suppose one chooses plane traveling waves $\Big\{ \frac{e^{i{\bf kr}}}{\sqrt{V}}\,| \ {\bf k} \neq 0 \Big\}$ as the bare-atom excited-state basis. Then the Hamiltonian in Eq.~(\ref{H-k}) acquires a diagonal form with the canonical Bogoliubov spectrum of eigen energies,
\begin{equation} \label{diagH-k}
\hat{H} = \sum_{{\bf k}\neq 0} E_{\bf k} \hat{\tilde{a}}_{\bf k}^\dagger \hat{\tilde{a}}_{\bf k},
\quad E_{\bf k} = \sqrt{\epsilon_{\bf k}^2 + 2 g N_0 \epsilon_{\bf k}}, 
\end{equation}
for the quasiparticle creation and annihilation operators $\hat{\tilde{\bf a}}^\dagger = \{\hat{\tilde{a}}_{\bf k}^\dagger |\, {\bf k} \neq 0 \}^\tp, \hat{\tilde{\bf a}} = \{ \hat{\tilde{a}}_{\bf k}|\, {\bf k} \neq 0 \}^\tp$ related to the corresponding bare-atom operators $\hat{\bf a}^\dagger = \{\hat{a}_{\bf k}^\dagger |\, {\bf k} \neq 0 \}^\tp, \ \hat{\bf a} = \{ \hat{a}_{\bf k}|\, {\bf k} \neq 0 \}^\tp$ as follows 
\begin{equation} \label{qp-k}
\hat{a}_{\bf k} = \bar{u}_{\bf k} \hat{\tilde{a}}_{\bf k} + \bar{v}_{\bf -k}^* \hat{\tilde{a}}_{\bf -k}^\dagger, \qquad
\hat{\tilde{a}}_{\bf k} = \bar{u}_{\bf k} \hat{a}_{\bf k} - \bar{v}_{\bf -k}^* \hat{a}_{\bf -k}^\dagger .
\end{equation}
Here we introduced the amplitudes $\bar{u}_{\bf k}$ and $\bar{v}_{\bf k}^*$,
\begin{equation} \label{eB-k}
\begin{split}
&\bar{u}_{\bf k} = \frac{\epsilon_{\bf k}+E_{\bf k}}{2\sqrt{\epsilon_{\bf k} E_{\bf k}}} \equiv \frac{1}{2} \left( \xi_{\bf k} + \frac{1}{\xi_{\bf k}} \right), 
\\
&\bar{v}_{\bf k}^* = \frac{\epsilon_{\bf k}-E_{\bf k}}{2\sqrt{\epsilon_{\bf k} E_{\bf k}}} \equiv -\frac{1}{2} \left( \xi_{\bf k} - \frac{1}{\xi_{\bf k}} \right); \quad \xi_{\bf k} = \sqrt{\frac{E_{\bf k}}{\epsilon_{\bf k}}},
\end{split}
\end{equation}
of the functions constituting the two-component wave function of those traveling-plane-wave quasiparticles
\begin{equation} \label{uv-k}
u_{\bf k} ({\bf r}) = \bar{u}_{\bf k} \ \frac{e^{i{\bf kr}}}{\sqrt{V}}, \qquad v_{\bf k}^* ({\bf r}) = \bar{v}_{\bf k}^* \ \frac{e^{-i{\bf kr}}}{\sqrt{V}} .
\end{equation}
In terms of such quasiparticles, the excited-atom field operator in Eq.~(\ref{field}) acquires the following explicit form
\begin{equation} \label{field operator-exp}
\begin{split}
\hat\psi_\ex({\bf r}) = \sum_{{\bf k}\neq 0} \Big( \, u_{\bf k}({\bf r}) \hat{\tilde{a}}_{\bf k} + v_{\bf k}^* ({\bf r}) \hat{\tilde{a}}_{\bf k}^\dagger \, \Big).
\end{split}
\end{equation}

Generally the quasiparticles are completely independent on each other. Thus, the correlations between the quasiparticle creation/annihilation operators $\hat{\tilde{\bf b}}^\dagger, \hat{\tilde{\bf b}}$, analogous to the bare-atom correlations in Eq.~(\ref{G=x}), are given by the $(2\times 2)$-block diagonal matrix of the thermal occupations of quasiparticles
\begin{equation} \label{DD}
   D = \begin{bmatrix}  \textrm{diag} \left\{(e^{E_l/T}-1)^{-1}\right\}  &  \mathbb{0} \\
    \mathbb{0}  &  \textrm{diag} \left\{(e^{E_l/T}-1)^{-1}\right\}   \end{bmatrix}.
\end{equation}

Employing the canonical Bose commutation relations $\hat{b}_l \hat{b}_{l'}^\dagger - \hat{b}_l^\dagger \hat{b}_{l'} = \delta_{l,l'}$ for bare atoms, it is easy to see that the covariance matrix (\ref{G=x}) of bare-atom creation/annihilation operators can be obtained from the matrix $D$ in a compact form,
\begin{equation} \label{GviaR}
    G = R \bigg( D + \frac{\mathbbm{1}}{2}\bigg) R^\dagger -\frac{\mathbbm{1}}{2} \ ,
\end{equation}
via the Bogoliubov transformation which relates the uncorrelated quasiparticles to the squeezed and interfering bare-atom excitations. It is described by the $(2\times 2)$-block symplectic matrix \cite{PRA2022} $\tilde{R}$ and its inverse matrix $R = \tilde{R}^{-1}$ as follows
\begin{equation} \label{generalBog}
\begin{split}
     &\bigg(  \begin{matrix}  {\,\bf \hat{\tilde{b}}^\dagger}   \\
                            {\bf \hat{\tilde{b}}}           \end{matrix} \bigg)
    = \ \tilde{R} \ \bigg(  \begin{matrix}  {\,\bf \hat{b}^\dagger} \\
                            {\bf \hat{b}}           \end{matrix} \bigg), \quad 
\tilde{R} = \bigg[  \begin{matrix}  A^\dagger  &  -B^\dagger     \\
                        -B^\tp  &  A^\tp        \end{matrix} \bigg]; \\
    &\bigg(  \begin{matrix}  {\,\bf \hat{b}^\dagger}   \\
                            {\bf \hat{b}}           \end{matrix} \bigg)
    = \ R \ \bigg(  \begin{matrix}  {\,\bf \hat{\tilde{b}}^\dagger} \\
                            {\bf \hat{\tilde{b}}}           \end{matrix} \bigg), \quad 
R = \bigg[  \begin{matrix}  A  &  B^*     \\
                        B  &  A^*        \end{matrix} \bigg], \qquad \tilde{R} = R^{-1} .                   
\end{split}
\end{equation}

Below we assume that some finite number $M$ of orthogonal excited states (i.e., atom wave functions) $\{ \phi_l|l=1,\ldots, M \}$ are preselected for sampling, that is, for a multi-detector measurement of the atom numbers, and constitute a basis of a finite-dimensional subspace of the single-particle Hilbert space $\mathcal{H}$. 
Their normal and anomalous correlations are given by the corresponding $(2M\times 2M)$-submatrix of the covariance matrix in Eq.~(\ref{G=x}). 
Suppose the preselected wave functions are coupled via the Bogoliubov Hamiltonian only between themselves. 
That is, their off-diagonal couplings $\Delta_{ll'}, \tilde{\Delta}_{ll'}$, and $\epsilon_{ll'}$ in Eq.~(\ref{HH}) with any wave functions outside the preselected subspace are zero or negligible. 
For example, in the uniform box trap the s-wave scattering in the Bogoliubov Hamiltonian couples just plane waves with opposite wave vectors ${\bf k}$ and $-{\bf k}$. 
Then the analysis gets easier. All bold-faced vectors (such as ${\bf \hat{b}}$) are reduced to $M$-dimensional vectors. 
All $(2\times 2)$-block matrices (including the Hamiltonian, $H$, Bogoliubov, $R$, quasiparticle occupation, $D$, covariance, $G$, and variable, $Z$, matrices) are reduced to the $(2M\times 2M)$-matrices containing corresponding $(M\times M)$-blocks such as $\Delta, \tilde{\Delta}, \epsilon, A, B$, etc. 
For the sake of simplicity, we'll denote any such matrix of a finite, reduced dimension $2M\times 2M$ or $M\times M$ by the same symbol that stands for its infinite-dimensional counterpart.

\section{The hafnian master theorem and sampling probabilities}

The hafnian master theorem recently found in \cite{PRA2022,LAA2022} provides the most convenient and powerful regular method for the analysis of the atomic, gaussian boson sampling and other problems associated with the $\sharp$P-hard computational complexity. 
The point is that it directly reduces a $\sharp$P-hard-for-computing quantity in question to a rigorously defined, canonical mathematical function --- a matrix hafnian (or its particular case --- a matrix permanent) computation of which belongs to the hardest, $\sharp$P-complete class of computational complexity. 
In accord with the famous Toda's theorem \cite{Toda1991,Basu2012}, it means that computing the hafnian and using it as an oracle is enough for polynomial-time reduction of every other $\sharp$P-complete or $\sharp$P-hard problem to an easy, polynomial-time computational problem. 

In our case the hafnian master theorem gives an explicit Fourier series (that could be viewed also as a Taylor expansion) of the characteristic function in Eq.~(\ref{CFdet}),
\begin{equation} \label{ThetaHMT}
 \Theta \big(\{ z_l \}\big) = \sum_{\{ n_l \}} \frac{ \haf \tilde{C}\big(\{ n_l \}\big)}{\sqrt{\det (\mathbbm{1}+G)}} \prod_{l} \frac{z_l^{n_l}}{n_l!} \ .
\end{equation}
It is expressed via the hafnian of an extended covariance-related $(2n\times 2n)$-matrix $\tilde{C}\big(\{ n_l \}\big)$, which has a dimension determined by the total atom number $n = \sum_l n_l$ in the sample of occupations of the preselected excited states $\{ \phi_l \}$ and is built from the covariance-related matrix 
\begin{equation} \label{C}
C = PG( \mathbbm{1} +G)^{-1}, \qquad P = \bigg[  \begin{matrix}  \mathbb{0}  &  \mathbbm{1}     \\
                \mathbbm{1} &  \mathbb{0}        \end{matrix} \bigg] .
\end{equation}
One has to replace the $l$-th and $(M+l)$-th rows with the $n_l$ copies of the $l$-th and $(M+l)$-th rows, respectively, and then the $l$-th and $(M+l)$-th columns with the $n_l$ copies of the $l$-th and $(M+l)$-th columns, respectively. If $n_l = 0$, then the corresponding rows and columns should be erased. 
The matrix $P$ just permutes the diagonal and off-diagonal blocks of the $(2\times 2)$-block matrix $G(\mathbbm{1}+G)^{-1}$ in a way appropriate for the hafnian. 

The concept of the matrix hafnian had been introduced in the quantum field theory \cite{Caianiello1953,Caianiello1973}. It expresses the Wick's, or Isserlis', theorem \cite{Wick1950,Barvinok2016} stating that the mean value, denoted below by angles, of the product of an even number, $2M$, of the centered Gaussian random variables $y_1, y_2, \ldots, y_{2M-1}, y_{2M}$ is equal to the hafnian of their covariance $2M\times 2M$ matrix $\bar{G}$, 
\begin{equation} \label{Wick}
\haf \bar{G} = \la y_1 y_2 \ldots y_{2M-1} y_{2M} \ra .
\end{equation}

Employing the identity 
\begin{equation}
\rho\big(\{ n_l \}\big) = \prod_l \frac{ \partial^{n_l}}{n_l! \, \partial z_l^{n_l}} \, \Theta \, \Big|_{\{ z_l=0 \}}\!, 
\label{pdfTaylor}
\end{equation}
following from the definition of the characteristic function,
we get an explicit analytical formula for the joint probabilities of the atom numbers in Eq.~(\ref{PDF}) via the hafnian as follows 
\begin{equation} \label{pdf=Hafnian}
\rho\big(\{ n_l \}\big) = \frac{ \haf \tilde{C}(\{ n_l \})}{\sqrt{\det (\mathbbm{1}+G)} \prod_l n_l!} \ . 
\end{equation}

A similar formula for the probabilities of output occupation numbers appears also in the theory of photonic boson sampling of Gaussian states in a linear interferometer \cite{Hamilton2017,HamiltonPRA2019}.

\section{Two ingredients of quantum supremacy and $\sharp$P-hardness: Squeezing and interference} 

The result in Eq.~(\ref{pdf=Hafnian}) explicitly shows that the $\sharp$P-hard complexity and, hence, potential quantum supremacy of atomic boson sampling from the $\sharp$P-hardness of computing the hafnian of the extended covariance-related matrix $\tilde{C}\big(\{ n_l \}\big)$ that depends only on the nontrivial Bogoliubov-transformation matrix $R$ and mean quasiparticle occupations constituting the diagonal matrix $D$ in Eq.~(\ref{DD}). Thus, the mystery of quantum supremacy is encoded in the structure of the Bogoliubov-transformation matrix in Eq.~(\ref{generalBog}) and can be revealed via its unique, irreducible Bloch-Messiah representation \cite{Braunstein2005,BlochMessiahPRA2016,Vogel2006,Huh2017,Huh2020MB} 
\begin{equation} \label{BlochMessiah}
\begin{split}
&\qquad \qquad \bigg(  \begin{matrix}  {\,\bf \hat{\tilde{b}}^\dagger}   \\
                            {\bf \hat{\tilde{b}}}           \end{matrix} \bigg)
= \tilde{R}_W \tilde{R}_r \tilde{R}_V \bigg(  \begin{matrix}  {\,\bf \hat{b}^\dagger}   \\
                            {\bf \hat{b}}           \end{matrix} \bigg); 
\\
&\tilde{R}_W =  \begin{bmatrix} W^*     &   \mathbb{0} \\
                                \mathbb{0} & W     \end{bmatrix}, \ 
\tilde{R}_r =   \begin{bmatrix} \cosh\,\Lambda_r    &   \sinh\,\Lambda_r \\
                            \sinh\,\Lambda_r    &   \cosh\,\Lambda_r    \end{bmatrix}, \
\tilde{R}_V =   \begin{bmatrix} V^*   &   \mathbb{0}   \\
                                \mathbb{0}   &   V \end{bmatrix}.
\end{split}
\end{equation}
It gives the blocks of the Bogoliubov-transformation matrix in Eq.~(\ref{generalBog}) in the form of a singular value decomposition,
\begin{equation}
A = V^\tp \cosh \Lambda_r  W^\tp,\qquad  B = -V^\dagger \sinh \Lambda_r W^\tp ,
\end{equation}
and involves two unitary matrices, $W$ and $V$, as well as the diagonal matrix of $M$ single-mode squeezing parameters
\begin{equation} \label{Lambda_r}
\Lambda_r = {\rm diag} \{ r_l |\, l=1,...,M \}, \quad r_l \geq 0.
\end{equation}

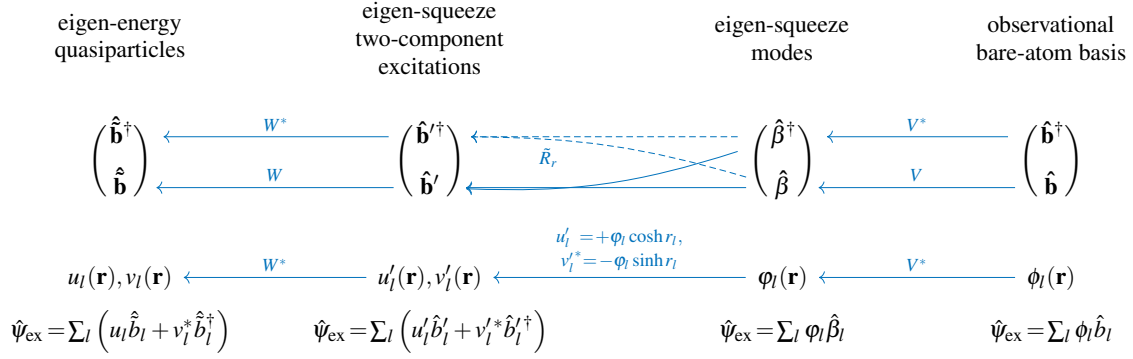
\begin{figure*} 
\begin{tikzcd}[row sep=1em,column sep=0em, minimum width=3em]
    \parbox{2.5cm}{\centering eigen-energy quasiparticles} &[2.5em]
    \parbox{2.5cm}{\centering eigen-squeeze two-component excitations} &[5em]
    \parbox{2.5cm}{\centering eigen-squeeze modes} &[2.5em]
    \parbox{2.5cm}{\centering observational bare-atom basis}
    \\
    {\raisebox{-1em}{$\Biggl($}\,\bf \hat{\tilde{b}}^\dagger\raisebox{-1em}{$\Biggr)$}} &
    \!\!{\raisebox{-1em}{$\Biggl($}\,\bf {\hat{b}'}{}^\dagger\raisebox{-1em}{$\Biggr)$}}\!\!
    \arrow[l, RoyalBlue, "W^*"'] &
    {\raisebox{-1em}{$\Biggl($}\,\bf \hat{\beta}^\dagger\raisebox{-1em}{$\Biggr)$}} 
    \arrow[ld, RoyalBlue, bend left=10, pos=0.32]
    \arrow[l,  RoyalBlue, dashed, yshift = 0ex]&
    {\raisebox{-1em}{$\Biggl($}\,\bf \hat{b}^\dagger\raisebox{-1em}{$\Biggr)$}}
    \arrow[l, RoyalBlue, "V^*"'] \\[-3em]
    {\bf \hat{\tilde{b}}} & 
    {\bf \hat{b}'}
    \arrow[l, RoyalBlue, "W"'] &
    {\bf \hat{\beta}}
    \arrow[lu, RoyalBlue, bend right=8, pos=0.6, yshift = 0ex, dashed, "\tilde{R}_r"]
    \arrow[l, RoyalBlue, yshift = 0ex]& 
    {\bf \hat{b}}
    \arrow[l, RoyalBlue, "V"']
    \\[+1em]
    u_l({\bf r}),v_l({\bf r}) & 
    u_l'({\bf r}),v'_l({\bf r}) \arrow[l,RoyalBlue,"W^*"']&
    \varphi_l({\bf r}) \arrow[l,RoyalBlue,"{\begin{smallmatrix}
                                    u_l' \,\ = \, +\varphi_l \cosh r_l, \\ 
                                    {v_l'}^{*} = \, -\varphi_l \sinh r_l  
                                \end{smallmatrix}}"']&
    \phi_l({\bf r}) \arrow[l,RoyalBlue,"V^*"']&
    \\[-1em]
    \hat\psi_\ex \!=\! \sum_l \Big(u_l \hat{\tilde{b}}_l + v_l^* \hat{\tilde{b}}^\dagger_l\Big)&
    \hat\psi_\ex \!=\! \sum_l \Big(u'_l \hat{b}'_l + v'_l{}^* {\hat{b}}'_l{}^\dagger\Big)&
    \hat\psi_\ex \!=\! \sum_l \varphi_l \hat\beta_l&
    \hat\psi_\ex \!=\! \sum_l \phi_l \hat{b}_l&
\end{tikzcd}
\caption{\label{diag} Schematic diagram of the Bloch-Messiah reduction in Eq.~(\ref{BlochMessiah}): Three irreducible steps of the Bogoliubov transformation (\ref{BlochMessiah}) of the creation/annihilation operators and wave functions from the observational bare-atom basis to the quasiparticle basis.}
\end{figure*}

\noindent The Bloch-Messiah reduction (\ref{BlochMessiah}) can be written in the form
\begin{equation}
\begin{split}
&\begin{bmatrix}  \cosh\,r^*          &   (-\sinh\,r^*) \,W^*W^\dagger    \\
                            (-\sinh\,r) \, WW^\tp     &   \cosh\,r    \end{bmatrix}
        \begin{pmatrix}  {\,\bf \hat{\tilde{b}}^\dagger}   \\
                            {\bf \hat{\tilde{b}}}           \end{pmatrix}
\\
&=
    \begin{bmatrix} (WV)^* & \mathbb{0}     \\ 
                    \mathbb{0} & W V     \end{bmatrix}
    \begin{pmatrix}  {\,\bf \hat{b}^\dagger}   \\
                            {\bf \hat{b}}      \end{pmatrix}
\end{split}
\end{equation}
that states that the effective evolution (resembling the Hamiltonian one, $e^{-i\hat{H}t}$) of the quasiparticle and bare-atom creation/annihilation operators under the action of the unitary multimode squeeze, $\hat{S}$, and rotation, $\hat{\Phi}$, operators, respectively, yields the same creation/annihilation operators: 
\begin{equation} \label{B=evolution}
\begin{split}
&\hat{S}^\dagger {\bf \hat{\tilde{b}}^\dagger} \hat{S} = \hat{\Phi}^\dagger {\bf \hat{b}^\dagger} \hat{\Phi} ,
\\
&\hat{S}^\dagger {\bf \hat{\tilde{b}}} \hat{S} = \hat{\Phi}^\dagger {\bf \hat{b}} \hat{\Phi} .
\end{split}
\end{equation} 
Such an interpretation, typical for quantum optics \cite{Braunstein2005,BlochMessiahPRA2016,Vogel2006,Huh2017,Huh2020MB,Ma1990}, employs the multimode squeeze and rotation operators,
\begin{equation} \label{S}
\hat{S} = \exp \bigg[ \frac{1}{2} \Big( {\bf \hat{\tilde{b}}^{\dagger\tp}} S \ {\bf \hat{\tilde{b}}^\dagger} - \, {\bf \hat{\tilde{b}}^\tp} S^\dagger {\bf \hat{\tilde{b}}} \Big) \bigg] ,
\end{equation}
\begin{equation} \label{RotationOperator}
\hat{\Phi} = \exp \big( i \, {\bf \hat{b}^{\dagger \tp}} \Phi \, {\bf \hat{b}} \big), 
\end{equation}
determined by a symmetric matrix $S = r e^{i\theta} = (W \Lambda_r W^\dagger) (-WW^\tp) = - W \Lambda_r W^\tp$ (built of the Hermitian matrices $r$ and $\theta$; the unitary matrix $e^{i\theta} = -WW^\tp$ is symmetric) and a Hermitian matrix $\Phi = \Phi^\dagger$ generating the unitary $e^{i\Phi} = WV$, respectively. 
The Hermitian factor, $r$, of the multimode squeeze matrix, $S$, is a positive semi-definite Hermitian $M\times M$ matrix whose diagonal representation $\Lambda_r$ is determined by the unitary $W$ as follows 
\begin{equation} \label{W-eigen-squeezing}
r = W \Lambda_r W^\dagger, \quad r \, {\bf w}_{l'} = r_{l'} {\bf w}_{l'} , \quad \big( W_{ll'} \big) = \big( ({\bf w}_{l'})_l \big),
\end{equation}
so that $\cosh r = W \cosh \Lambda_r W^\dagger$, $\sinh r = W \sinh \Lambda_r W^\dagger$. 

The unitaries $W$ and $V$ are chosen to satisfy the so-called rotation condition emphasized in \cite{BlochMessiahPRA2016} in view of a possible nonuniqueness of the singular value decomposition, particularly in the presence of degenerate singular values. 
The singular vectors $\{ {\bf w}_l|\,l=1,...,M \}$ and singular values $\{ r_{l} \geq 0|\,l=1,...,M \}$ are the eigenvectors (comprising the unitary $W$ as columns) and the eigenvalues of the Hermitian factor $r$ of the squeeze matrix $S$, respectively. The squeezing parameters, or eigenvalues, $\{ r_l \}$ constitute the unique, irreducible resource of the many-body interacting system and do not depend on the choice of bases or unitaries. 

Eq.~(\ref{BlochMessiah}) describes the overall Bogoliubov transformation as a sequence of three $(2\times 2)$-block transformations $\tilde{R}_W \tilde{R}_r \tilde{R}_V$ from the creation/annihilation operators in the observational basis of the excited atom states preselected for sampling measurement to the creation/annihilation operators of completely independent quasiparticles which correspond to the diagonalized Hamiltonian in Eq.~(\ref{QPH}) and stay in totally separable, disentangled equilibrium states. Their combined state is described by the density matrix, or statistical operator, 
\begin{equation} \label{qprho} 
\hat{\rho} = \prod_l e^{-E_l \hat{\tilde{b}}_l^\dagger \hat{\tilde{b}}_l/T} \Big/ \tr \Big\{ e^{-E_l \hat{\tilde{b}}_l^\dagger \hat{\tilde{b}}_l/T} \Big\},
\end{equation}
expressed in terms of the eigen-energy quasiparticle creation and annihilation operators.

The three $2\times 2$ blocks of the Bogoliubov reduction in Eq.~(\ref{BlochMessiah}) could be thought of as the matrices transforming the annihilation and creation operators or the corresponding wave functions in Eq.~(\ref{field:ba-qp}). This is explained below by means of the schematic diagram in Fig.~\ref{diag} and Eqs.~(\ref{unitaryMixingV})-(\ref{eigen-squeeze}). 

The first part of the Bogoliubov transformation, $\tilde{R}_V$, corresponds to the unitary rotation $V^*$, 
\begin{equation} \label{unitaryMixingV}
\varphi_l({\bf r}) = \sum_{l'} V^*_{ll'} \phi_{l'}({\bf r}),
\quad
\hat\beta_l = \sum_{l'} V_{ll'} \hat{b}_{l'}, 
\quad \hat\psi_\ex({\bf r}) = \sum_l \varphi_l({\bf r}) \hat\beta_l ,
\end{equation}
of the basis of the bare-atom excited states $\{\phi_l|\, l=1,...,M\}$, prescribed for measurement of atom numbers by means of multi-detector imaging in the process of sampling, into the basis of the eigen-squeeze modes $\{\varphi_l\}$ associated with the eigenvectors $\{{\bf w}_l\}$ of the Hermitian factor of the squeeze matrix $r$ in Eq.~(\ref{W-eigen-squeezing}) and explicitly expressed below in Eq.~(\ref{eigen-squeeze}) via the quasiparticle wave functions. The corresponding transformation from annihilation operators of the observable states, $\{ \hat{b}_l \}$, to the annihilation operators of the eigen-squeeze modes, $\{ \hat{\beta}_l \}$, is performed by the unitary $V$ and does not involve creation operators. 

The second part of the Bogoliubov transformation, $\tilde{R}_r$, is associated with the presence in the Hamiltonian the terms beyond the resonant-wave approximation (non-RWA terms), $\hat{b}_l^\dagger \hat{b}_{l'}^\dagger$ and $\hat{b}_l \hat{b}_{l'}$, which create or annihilate a pair of excited atoms. It converts the state of the system into the squeezed state in which each single-particle eigen-squeeze mode $\varphi_l({\bf r})$ constitutes the same spatial profile for both wave-function components $u'_l ({\bf r}),  v'^*_l ({\bf r})$ of the $l$-th eigen-squeeze two-component excitation which acquires the corresponding nontrivial squeezing parameter $r_l \geq 0$. Note that, according to Eq.~(\ref{eigen-squeeze}), those two components have different amplitudes, $+\cosh r_l$ and $-\sinh r_l$. 

The third part of the Bogoliubov transformation, $\tilde{R}_W$, relates the creation, $\{\hat{b}'_{l'}{}^\dagger \}$, and annihilation, $\{\hat{b}'_{l'} \}$, operators of the eigen-squeeze two-component excitations, formed on the second step, to the eigen-energy-quasiparticle creation, $\{\hat{\tilde{b}}_l^\dagger \}$, and annihilation, $\{\hat{\tilde{b}}_l \}$, operators:
\begin{equation} \label{eigen-squeezeOperators}
    \hat{b}'_{l'}{}^\dagger = \sum_l W_{ll'} \hat{\tilde{b}}_l^\dagger, 
    \qquad 
    \hat{b}'_{l'} = \sum_l W_{ll'}^* \hat{\tilde{b}}_l .
\end{equation}
According to the form of the quasiparticle field operator in Eq.~(\ref{field:ba-qp}), it means two simultaneous unitary rotations (under the action of one and the same unitary matrix $W^\dagger$) of the single-squeeze modes $\{\varphi_l\}$ into the wave functions $\{u_l|\, l=1,...,M\}$ and $\{v_l^*|\, l=1,...,M\}$, which according to Eq.~(\ref{field:ba-qp}) constitute the bases of the first and second components of the two-component functional space of the quasiparticle wave functions $(u_l({\bf r}), v_l^*({\bf r}))$, 
\begin{equation} \label{eigen-squeeze-to-qp}
\begin{split}
    &u_l ({\bf r}) = +\sum_{l'} W_{ll'}^* (\cosh r_{l'}) \varphi_{l'} ({\bf r}), 
    \\  
    &v_l^* ({\bf r}) = -\sum_{l'} W_{ll'} (\sinh r_{l'}) \varphi_{l'} ({\bf r}).
\end{split}    
\end{equation}
Since the quasiparticle wave functions $\{u_l\}$ and $\{v_l^*\}$ are fully predetermined, fixed by the coupling parameters (interactions) in the Hamiltonian (\ref{HH}) (of course, up to a possible degeneracy), the unitary matrix $W = (W^{\dagger})^{-1}$ determines the unique (up to a possible degeneracy) eigen-squeeze modes $\{\varphi_{l'}({\bf r})\}$ which are the same for both components of the quasiparticle functional space as per equations inverse to Eqs.~(\ref{eigen-squeeze-to-qp}):
\begin{equation} \label{eigen-squeeze}
\begin{split}
&+(\cosh r_{l'}) \varphi_{l'}({\bf r}) = u'_{l'} ({\bf r}) = \sum_l W_{ll'} \, u_l ({\bf r}), \\ 
&-(\sinh r_{l'}) \varphi_{l'} ({\bf r}) = v'^*_{l'} ({\bf r}) = \sum_l W^*_{ll'} \, v_l^* ({\bf r}).
    \end{split}
\end{equation}
Each single-squeeze two-component excitation originating from the mode $\varphi_{l'}({\bf r})$ owns the single-mode squeezing parameter $r_{l'} \geq 0$ and is not subject to inter-mode squeezing with other eigen-squeeze two-component excitations. 

The existence of such a unique unitary $W$, simultaneously converting the basis wave functions $u_l({\bf r})$ and $v_l^*({\bf r})$ of both components of the two-component functional space of quasiparticles into the basis wave functions $u'_{l'}({\bf r}) = +(\cosh r_{l'}) \varphi_{l'}({\bf r})$ and $v'^*_{l'}({\bf r}) = -(\sinh r_{l'}) \varphi_{l'} ({\bf r})$ which are equal to the same eigen-squeeze mode $\varphi_{l'}({\bf r})$ just multiplied by different constant factors $\cosh r_l$ and $-\sinh r_l$, respectively, is a nontrivial and important property of the Bogoliubov transformation. 
It is a consequence of the symplectic property \cite{PRA2022}, 
\begin{equation} \label{symplectic}
R \ \Omega \ R^\tp = \Omega, \qquad \Omega = \bigg[  \begin{matrix}  \mathbb{0}    & \mathbbm{1}      \\
                - \mathbbm{1} & \mathbb{0}     \end{matrix} \bigg],
\end{equation}
of the Bogoliubov transformation, that is, canonical Bose commutation relations, and highlights the fact that the two components of the quasiparticle eigen function $(u_l({\bf r}), v_l^*({\bf r}))$ are not independent, but, on the contrary, are deeply inter-correlated. 

Both the first, $\tilde{R}_V$, and the third, $\tilde{R}_W$, factors of the Bogoliubov transformation (\ref{BlochMessiah}) introduce unitary interference, that is entanglement, into the quantum many-body state  (statistical operator $\hat{\rho}$) of $M$ excited atom modes chosen for atomic boson sampling. 
They are controlled by two different means. 
The trapping potential and parameters of the Hamiltonian in Eq.~(\ref{HH}), that is, couplings $\tilde{\Delta}_{ll'}, \Delta_{ll'}, \epsilon_{ll'}$ and atom interactions, control both unitaries $W$ and $V$ since they determine the composition of the eigen-squeeze modes both with respect to the quasiparticle wave functions and the bare-atom wave functions. 
The unitary $V$ of the first factor is additionally controlled via a choice of different excited atom states by means of reconfiguring the multiple detectors for atomic sampling. 
In the particular case of a box trap with a uniform condensate, discussed in the present paper, the third Bogoliubov transformation $\tilde{R}_W$ is almost completely fixed. 
Yet, the first Bogoliubov transformation $\tilde{R}_V$ provides an access to practically arbitrary unitary matrices $V$, controllable in a wide range of their parameters. 
This should be enough for ensuring $\sharp$P-hardness of computing the hafnian in Eq.~(\ref{pdf=Hafnian}) on average. 
Such complexity on average \cite{Harrow2017,HamiltonPRA2019} is important for a possibility of demonstrating quantum supremacy of atomic boson sampling.  

Thus, the Bloch-Messiah reduction in Eq.~(\ref{BlochMessiah}) unambiguously specifies two preferred bases: (i) the basis of the quasiparticle operators in Eq.~(\ref{field:ba-qp}) ensuring the diagonal form of the Hamiltonian (\ref{QPH}) and (ii) the basis of the eigen-squeeze single-particle excited states in Eq.~(\ref{eigen-squeeze}) diagonalizing the Hermitian factor of the multimode squeeze matrix, Eq.~(\ref{W-eigen-squeezing}). Moreover, the Bogoliubov transformation in the Bloch-Messiah representation explicitly relates both above-mentioned bases to the observational basis of excited bare-atom states $\{ \phi_l|\, l=1,...,M \}$ which can be arbitrarily selected by a reconfiguration of atom detectors. 

The interference between wave functions of those bases, appearing in the statistical operator (state) of the atomic many-body (multimode) system due to the unitaries $W, V$ and leading to the entanglement of different Bose bare-atom modes as opposed to the separability of quasiparticle states, constitutes the first of the two ingredients of the computational $\sharp$P-hardness and potential quantum supremacy of the atomic boson sampling in equilibrium. 

The second ingredient is the squeezing of the equilibrium state of the excited-atom modes originating from the reduced, canonical Bogoliubov transformation $\tilde{R}_r$ determined exclusively by the eigenvalues $\{ r_l \}$ of the Hermitian factor of the multimode squeeze matrix $r$. 
If it was given by the identity matrix with all single-squeezing parameters equal zero, $\{ r_l =0 \}$, then the hafnian in Eq.~(\ref{pdf=Hafnian}) would be reduced to the permanent of a positive matrix, which according to \cite{Aaronson2013,LundPRL2014,Lund2015} could be approximated by the Stockmeyer’s approximating algorithm \cite{Stockmeyer} in a computational-complexity class simpler than $\sharp P$.
A physical mechanism of squeezing can be seen from Eq.~(\ref{field:ba-qp}) for the atomic field operator. 
Any quantum or thermal fluctuation associated with disappearance of an atom at a point ${\bf r}$ implies annihilation of a superposition of quasiparticles with the amplitudes given by the first component of the quasiparticle wave functions $\{ u_l({\bf r})|l=1,\ldots,M \}$ and simultaneous creation of a superposition of quasiparticles with the amplitudes given by the second component of the quasiparticle wave functions $\{ v_l^*({\bf r})|\, l=1,...,M \}$. 

In the observational basis coinciding with the eigen-squeeze modes $\{\varphi_l\}$, when $\tilde{R}_V = \mathbbm{1}$, the covariance matrix (\ref{GviaR}) acquires a unique irreducible form $G = G_Q + G_T$ of a sum of the pure quantum (independent on temperature and associated with a quantum depletion of the condensate due to interatomic interaction) and complimentary thermal correlations between creation/annihilation operators $\hat{\beta}_l^\dagger, \hat{\beta}_l$ of the eigen-squeeze modes: 
\begin{equation} \label{GQ} 
G_Q = \frac{RR^\dagger - \mathbbm{1}}{2} = 
    \left[  \begin{matrix}\sinh^2\Lambda_r & -\sinh\Lambda_r \, \cosh\Lambda_r \\
    -\sinh \Lambda_r \, \cosh \Lambda_r & \sinh^2\Lambda_r    \end{matrix} \right] ,
\end{equation}
\begin{equation} \label{GT}
\begin{split}
G_T &= RDR^\dagger  \\
&=\left[  \begin{matrix}  \cosh \Lambda_r &   -\sinh \Lambda_r     \\
                        -\sinh \Lambda_r &   \cosh \Lambda_r     \end{matrix} \right]   
            \left[  \begin{matrix}  q^* &   \mathbb{0}  \\
                                    \mathbb{0}   &   q  \end{matrix} \right]
\left[  \begin{matrix}  \cosh \Lambda_r &   -\sinh \Lambda_r     \\
                        -\sinh \Lambda_r &   \cosh \Lambda_r \end{matrix} \right].
\end{split}
\end{equation}
Here $q = W^\dagger \, \text{diag}\{(e^{E_j/T}-1)^{-1}| j=1,\ldots,M\} \, W$ is the $M\times M$ matrix defined by thermal occupations of quasiparticles.
These irreducible contributions are determined exclusively by the two intrinsic entities of the BEC gas, the eigen-squeeze modes and the eigen-energy quasiparticles, and are not subjected to an arbitrariness of choosing any observational basis. 
In an arbitrary observational basis $\{\phi_l\}$ the covariance matrix appears as the corresponding unitary transform of the irreducible one as per unitary rotation in Eq.~(\ref{unitaryMixingV}),
\begin{equation} \label{CMcanon}
    G = \left[  \begin{matrix}  V^\tp & \mathbb{0}   \\
         \mathbb{0} &   V^\dagger     \end{matrix} \right]      
            ( G_Q + G_T )
    \left[  \begin{matrix}  V^* & \mathbb{0}   \\
                            \mathbb{0}  & V     \end{matrix} \right].
\end{equation}

Below we illustrate and discuss the interplay and manifestations of two aforestated ingredients of quantum supremacy in atomic boson sampling for different choices of the observational basis of excited bare-atom states $\{ \phi_l|\, l=1,...,M \}$. 

\section{Separability of atomic boson sampling in the basis of single-mode-squeezed standing plane waves: Atom-number statistics via the hafnian and Legendre polynomials}

A minimal complexity of joint atom-number probabilities in the atomic boson sampling is achieved if one chooses the eigen-squeeze modes in Eq.~(\ref{eigen-squeeze}) as the bare-atom excited states for atom-number measurements by the multi-detector imaging system. For the considered model of the uniform condensate in a box trap with the periodic boundary conditions, the wave functions of the eigen-energy quasiparticles could be chosen in such a way that they coincide with the eigen-squeeze modes. Let's choose them to be $\sin({\bf kr})$ and $\cos({\bf kr})$ spatial modes with a wave vector ${\bf k}$. In fact, any orthogonal (but not any unitary) transformation of a pair of energy-wise degenerate eigen-squeeze excited states would lead to the same atom-number sampling statistics.

Thus, the excited atoms are described by the field operator 
\begin{equation} \label{field-sc}
    \hat\psi_\ex({\bf r}) =  \sqrt{\frac{2}{V}} \ \sum_{{\bf k}\neq 0}{}^{'} \ \ \big[ \sin ({\bf k r}) \ \hat{s}_{\bf k} 
                    + \ \cos ({\bf kr}) \ \hat{c}_{\bf k} \big] ,
\end{equation}
written via operators $\hat{s}_{\bf k}, \hat{c}_{\bf k}$ annihilating a bare atom in the corresponding sinusoidal states with a wave vector ${\bf k} = \frac{2\pi{\bf j}}{L}, \ {\bf j} = \{j_x,j_y,j_z\} \in \mathbbm{N}^3$. Hereinafter, an apostrophe in the symbol of the sum $\sum^{'}_{{\bf k}\neq 0}$ means summation over nonzero wave vectors modulo multiplication by $-1$, that is, the integer vector ${\bf j} \equiv \{j_x,j_y,j_z\}$ is running over a half of a three-dimensional (3D) integer space, $\mathbbm{Z}^3_+ = \big\{ {\bf j} \ \big| \ j_z \ge 0 \ \& \ j_x^2+j_y^2+j_z^2 \ne 0 \big\}$, with its origin excluded. Such an enumeration is convenient for the subsequent analysis of the particular cases of standing plane waves, corresponding to the opposite wave vectors ${\bf k} = 2\pi {\bf j}/L$ and $-{\bf k} = -2\pi {\bf j}/L$, because it allows one to explicitly take into account the fact of a two-fold degeneracy of their energies in the box trap. In other words, for each pair of terms corresponding to the opposite nonzero wave vectors ${\bf k}$ and ${\bf -k}$ only one of the terms belongs to the sum. (Note that the sum in Eqs.~(\ref{field}), (\ref{field-k}) does not include the apostrophe.)

In such a sinusoidal, standing-wave basis of excited states, 
\begin{equation} \label{sc}
\{ \phi_l \} = \bigg\{ \frac{\sqrt{2}\sin({\bf kr})}{\sqrt{V}}, \frac{\sqrt{2}\cos({\bf kr})}{\sqrt{V}} \bigg|\, {\bf k} = \frac{2\pi{\bf j}}{L}, \ {\bf j} \in \mathbbm{Z}^3_+ \bigg\},
\end{equation}
the Hamiltonian (\ref{HH}) splits into two independent parts involving separate operators related to $\sin({\bf kr})$ or $\cos({\bf kr})$ modes, 
\begin{equation} \label{H-sc}
\begin{split}
\hat{H} =& \sum_{{\bf k,k'}\neq 0}{}^{'} \ \ \bigg[ \big(\epsilon_{\bf kk'} + \Delta_{\bf kk'}\big) \hat{s}_{\bf k}^\dagger \hat{s}_{\bf k} + \frac{1}{2} \tilde{\Delta}_{\bf kk'} \big(\hat{s}_{\bf k}^\dagger \hat{s}_{\bf k'}^\dagger + \hat{s}_{\bf k}\hat{s}_{\bf k'}\big) 
\\ 
&+ \big(\epsilon_{\bf kk'} + \Delta_{\bf kk'}\big) \hat{c}_{\bf k}^\dagger \hat{c}_{\bf k} + \frac{1}{2} \tilde{\Delta}_{\bf kk'} \big(\hat{c}_{\bf k}^\dagger \hat{c}_{\bf k'}^\dagger + \hat{c}_{\bf k} \hat{c}_{\bf k'}\big)  \bigg].
\end{split}
\end{equation}
It is in contrast to the Hamiltonian in Eq.~(\ref{H-k}) in which the modes $e^{i{\bf kr}}$ and $e^{-i{\bf kr}}$ of the traveling-plane-wave basis $\big\{ \frac{e^{i{\bf kr}}}{\sqrt{V}}\, \big| \ {\bf k} \neq 0 \big\}$ were coupled to each other. 
The Hamiltonian in Eq.~(\ref{H-sc}), compared to the Hamiltonian in Eq.~(\ref{H-k}), has the same bare-atom energies $\epsilon_{\bf k}$ and normal overlapping integrals $\Delta_{\bf kk'}$, but different, now diagonal anomalous overlapping integrals
\begin{equation} \label{edd}
\epsilon_{{\bf k}{\bf k'}} = \epsilon_{\bf k} \delta_{{\bf k},{\bf k'}}, \qquad \Delta_{{\bf k}{\bf k}'} = \tilde{\Delta}_{{\bf k}{\bf k}'} = gN_0 \delta_{{\bf k},{\bf k}'}. 
\end{equation}
In this case, the quasiparticle annihilation operators $\{\hat{\tilde{c}}_{\bf k}, \hat{\tilde{s}}_{\bf k}\}$ are given by a similar Bogoliubov transformation via the corresponding bare-particle annihilation and creation operators:
\begin{equation}
\begin{split}
&\hat{s}_{\bf k} = \bar{u}_{\bf k} \hat{\tilde{s}}_{\bf k} + \bar{v}_{\bf k} \hat{\tilde{s}}_{\bf k}^\dagger, \qquad \hat{c}_{\bf k} = \bar{u}_{\bf k} \hat{\tilde{c}}_{\bf k} + \bar{v}_{\bf k} \hat{\tilde{c}}_{\bf k}^\dagger;
    \\
&\hat{\tilde{s}}_{\bf k} = \bar{u}_{\bf k} \hat{s}_{\bf k} - \bar{v}_{\bf k} \hat{s}_{\bf k}^\dagger, \qquad \hat{\tilde{c}}_{\bf k} = \bar{u}_{\bf k} \hat{c}_{\bf k} - \bar{v}_{\bf k} \hat{c}_{\bf k}^\dagger.
\end{split}
\end{equation}
The quasiparticle operators diagonalize the Hamiltonian (\ref{H-sc}), 
\begin{equation} \label{eigen-squeeze energies}
\hat{H} = \sum_{{\bf k}\neq 0}\,^{'} \ \ E_{\bf k} \left(\hat{\tilde{s}}_{\bf k}^\dagger \hat{\tilde{s}}_{\bf k} + \hat{\tilde{c}}_{\bf k}^\dagger \hat{\tilde{c}}_{\bf k}\right), \quad E_{\bf k} = \sqrt{\epsilon_{\bf k}^2 + 2 g N_0 \epsilon_{\bf k}} \ \ ,
\end{equation}
and turn the field operator (\ref{field-sc}), annihilating bare atoms, into a sum of both annihilation and creation quasiparticle operators:
\begin{equation} \label{field operator-sin}
\hat\psi_\ex({\bf r}) = \sum_{{\bf k}\neq 0}\,^{'} \ \ \frac{\sin({\bf kr}) (\tilde{u}_{\bf k} \hat{\tilde{s}}_{\bf k} + \tilde{v}_{\bf k} \hat{\tilde{s}}_{\bf k}^\dagger) +
\cos({\bf kr}) (\tilde{u}_{\bf k} \hat{\tilde{c}}_{\bf k} + \tilde{v}_{\bf k} \hat{\tilde{c}}_{\bf k}^\dagger)}{\sqrt{V/2}}. 
\end{equation}
In both exponential and sinusoidal, bases, the quasiparticle energies $E_{\bf k}$ and factors $\bar{u}_{\bf k}, \bar{v}_{\bf k}$ are the same, given in Eqs.~(\ref{diagH-k}), (\ref{eB-k}). The energy $E_{\bf k}$ is larger than the bare-atom energy $\epsilon_{\bf k}$.

Thus, in the observational basis of the eigen-squeeze modes (\ref{sc}) the sampling joint probability distribution factorizes into the product of probabilities for single $\sin({\bf kr})$ or $\cos({\bf kr})$ modes. Each of such single-mode-squeezed probabilities can be calculated analytically by means of the general solution in Eq.~(\ref{pdf=Hafnian}). We just need to appreciate the fact that in this case the general Bloch-Messiah representation of the Bogoliubov transformation in Eq.~(\ref{BlochMessiah}) is reduced to the simple independent blocks with trivial unitary parts $\tilde{R}_W = \tilde{R}_V = \mathbbm{1}$, that is
\begin{equation} \label{scBog}
\begin{split}
&\left(  \begin{matrix} \hat{\tilde{s}}_{\bf k}^\dagger\\
        \hat{\tilde{s}}_{\bf k}\\
            \end{matrix} \right)
=  \begin{bmatrix}  \cosh r_{\bf k} & \sinh r_{\bf k} \\
                            \sinh r_{\bf k} & \cosh r_{\bf k} \\
            \end{bmatrix}
    \left(  \begin{matrix}  \hat{s}_{\bf k}^\dagger\\
            \hat{s}_{\bf k}
            \end{matrix} \right),
\\ 
&\begin{pmatrix} \hat{\tilde{c}}_{\bf k}^\dagger \\                           \hat{\tilde{c}}_{\bf k}\\
            \end{pmatrix} =
    \begin{bmatrix}  \cosh r_{\bf k} & \sinh r_{\bf k} \\
                            \sinh r_{\bf k} & \cosh r_{\bf k} \\
            \end{bmatrix}
    \begin{pmatrix}  \hat{c}_{\bf k}^\dagger\\
            \hat{c}_{\bf k}
            \end{pmatrix};
\quad 
r_{\bf k} = \frac{1}{2}\ln \frac{E_{\bf k}}{\epsilon_{\bf k}}.
\end{split}
\end{equation}
This is the Bogoliubov single-mode squeezing in the simplest, pure form with a single-mode squeezing parameter $r_{\bf k} \geq 0$. 

\begin{figure} 
\includegraphics{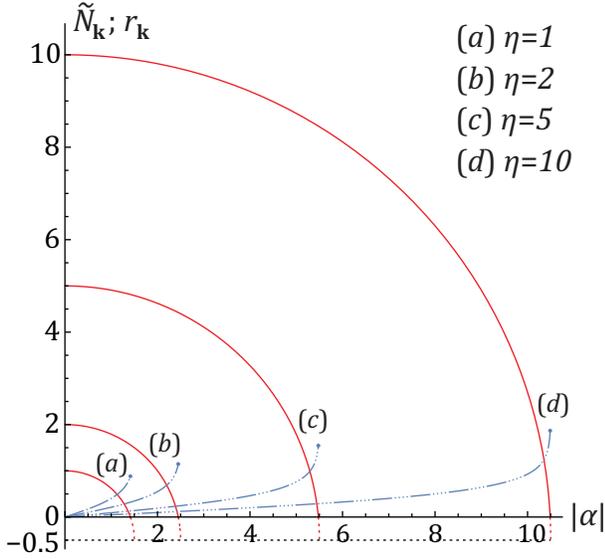}
\caption{   \label{fig:parameters}
An average quasiparticle occupation, $\tilde{N}_{\bf k}$, (red solid curves) and the single-mode squeezing parameter, $r_{\bf k}$, (blue dashed curves) vs the absolute value of the anomalous correlator $|\la \hat{s}_{\bf k}\hat{s}_{\bf k}\ra| \equiv |\alpha|$ for different fixed values of the normal correlator $\la \hat{s}_{\bf k}^\dagger\hat{s}_{\bf k}\ra \equiv \eta$.
Red solid curves representing the average quasiparticle occupation are, in fact, the arcs of a circle with a radius $\eta + 1/2$ centered at $(0,-1/2)$.
}
\end{figure}

The corresponding single-mode covariance matrix is given by Eqs.~(\ref{GQ}), (\ref{GT}) as follows
\begin{equation} \label{Gk}
\begin{split}
    &G^{({\bf k})} =  \left(\tilde{N}_{\bf k} + \frac{1}{2}\right) 
    \left[  \begin{matrix}  \cosh 2r_{\bf k}    &   -\sinh 2r_{\bf k} \\
                            -\sinh 2r_{\bf k}   &   \cosh 2r_{\bf k}     \end{matrix}    \right]    
    -\frac{\mathbbm{1}}{2}
     = \left[  \begin{matrix}   \eta    &   \alpha  \\
                                \alpha  &   \eta    \end{matrix}    \right];
    \\
    &\tilde{N}_{\bf k} = 1 \big/ \big( e^{E_{\bf k}/T} - 1 \big),
    \\
    &\eta \equiv \langle \hat{s}_{\bf k}^\dagger \hat{s}_{\bf k} \rangle  = \left(\tilde{N}_{\bf k} + 1/2\right)\cosh 2r_{\bf k} - 1/2,
    \\
&\alpha \equiv \langle \hat{s}_{\bf k} \hat{s}_{\bf k} \rangle = -\left(\tilde{N}_{\bf k} + 1/2\right)\sinh 2r_{\bf k}; \quad |\alpha| \le \sqrt{\eta(1+\eta)}.
\end{split}
\end{equation}
Here $\tilde{N}_{\bf k}, \eta$ and $\alpha$ stand for the average number of quasiparticles, normal correlator  and anomalous correlator of the single sinusoidal mode, respectively. 
The anomalous correlator is negative for the chosen phases of the basis wave functions and its absolute value is bounded from above by the inequality in Eq.~(\ref{Gk}) which is equivalent to the obvious inequality $\tilde{N}_{\bf k} \ge 0$. The maximum value of the anomalous correlator, $\max |\alpha| = \sqrt{\eta(1+\eta)}$, is achieved when $\tilde{N}_{\bf k} = 0$, that is, when the quasiparticles are in the vacuum state.
The relations between the average quasiparticle occupation $\tilde{N}_{\bf k}$, the single-mode squeezing parameter $r_{\bf k}$ and the normal and anomalous correlators $\eta$ and $\alpha$ are illustrated in Fig.~\ref{fig:parameters}.
Any fixed value of the normal correlator which is equal to the mean number of atoms in the eigen-squeeze mode, $\eta = \la n \ra$, corresponds to a circle on the plane $\big(|\alpha|, \tilde{N}_{\bf k}, \big)$ with the origin at the point $(0,-1/2)$, i.e.,
$(\tilde{N}_{\bf k}+1/2)^2 + |\alpha|^2 = (\eta+1/2)^2$.

The eigenvalues of this covariance matrix depend on the single-mode squeezing parameter $r_{\bf k}$ and the mean quasiparticle occupation $\tilde{N}_{\bf k}$, 
\begin{equation} \label{lambda_12}
    \lambda_{1,2} = \eta \pm |\alpha| = \tilde N_{\bf k}e^{\pm r_{\bf k}} + \big(e^{\pm r_{\bf k}}-1\big)/2.
\end{equation}

The atom-number probability distribution in a sinusoidal mode is determined, as per Eq.~(\ref{C}), by the covariance-related matrix $C = P G^{({\bf k})} \big(\mathbbm{1}+G^{({\bf k})}\big)^{-1}$ which has the following explicit form:
\begin{equation} \label{C-k}
\begin{split}
    &C  \equiv   \left[  \begin{matrix} \alpha'     &   \eta'   \\
                                        \eta'       &   \alpha'   \end{matrix} \right];
    \ \
    \eta' = \frac{\eta^2 + \eta - |\alpha|^2}{(\eta+1)^2-|\alpha|^2},
    \ \ \!
    \alpha' = \frac{\alpha}{(\eta+1)^2-|\alpha|^2},
    \\
    &\det \big(\mathbbm{1}+G^{({\bf k})}\big) = (\eta+1)^2-|\alpha|^2 = \frac{e^{2E_{\bf k}/T} - \tanh^2 r_{\bf k}}{(e^{E_{\bf k}/T}-1)^2 \cosh^2 r_{\bf k}}.
\end{split} 
\end{equation}
It is convenient to denote the entries $\alpha', \eta'$ of the matrix $C = P G^{({\bf k})} \big(\mathbbm{1}+G^{({\bf k})}\big)^{-1}$ by adding apostrophe to the symbols $\alpha, \eta$ denoting the entries of the covariance matrix $G^{({\bf k})}$. 
The eigenvalues $\lambda'_{1,2}$ of the renormalized covariance matrix $G^{({\bf k})} \big(\mathbbm{1}+G^{({\bf k})}\big)^{-1}$ look similar to the eigenvalues $\lambda_{1,2}$ of the covariance matrix $G^{({\bf k})}$, namely,
\begin{equation}    \label{lambda'}
    \lambda'_{1,2} = \eta'\pm|\alpha'| = \frac{\eta\pm|\alpha|}{1+\eta\pm|\alpha|} = \frac{1 \pm e^{E_{\bf k}/T}\tanh r_{\bf k}}{e^{E_{\bf k}/T} \pm \tanh r_{\bf k}}.
\end{equation}
Note that the maximal value of the absolute value of the anomalous correlator, $\max |\alpha| = \sqrt{\eta (1 + \eta)}$, is achieved when $\eta'=0$.

Now we calculate the atom-number probabilities. 
The result is immediately given by the hafnian master theorem (\ref{pdf=Hafnian}),
\begin{equation} \label{rho_n}
    \rho_n  = \frac{\haf \tilde{C}(n)}{n!\,\sqrt{\det(\mathbbm{1}+G^{({\bf k})})}},
    \quad
    \tilde{C}(n)
    =    \left[  \begin{matrix} \alpha'\, J_{n\times n}   & 
                                                    \eta'\, J_{n\times n} 
                                \\
                                \eta'\, J_{n\times n}       & 
                                                    \alpha'\, J_{n\times n} \\
                            \end{matrix} \right],
\end{equation}
where the extended covariance-related matrix $\tilde{C}(n)$ involves the $n\times n$ matrix $J_{n\times n}$ with all entries equal unity.

In the absence of anomalous correlations, when $\alpha' = 0$, the eigen-squeeze modes are in a non-squeezed thermal state and the hafnian in Eq.~(\ref{rho_n}) is reduced to the known permanent of the unity matrix $J_{n\times n}$, that is, $\haf \tilde{C}(n) = \per \big(\eta' J_{n\times n} \big) = n! (\eta')^n$. 
In this case the atom-number probability distribution is reduced to the simple exponential law
\begin{equation} \label{rho_n=exp}
\rho_n = \frac{1}{\tilde{N}_{\bf k} + 1} \left(\frac{\tilde{N}_{\bf k}}{\tilde{N}_{\bf k} + 1} \right)^{n}, \qquad \alpha' = 0,
\end{equation}
typical for a counting statistics of a mode in the thermal state \cite{Weedbrook2012,Barnett1996,Vogel2006}.  

The presence of a nonzero anomalous correlator $\alpha'$ introduces squeezing and makes this distribution nontrivial. Fortunately, the hafnian in Eq.~(\ref{rho_n}) in the case of an arbitrary $\alpha'$ is easy to calculate via the known recursive relation for hafnians \cite{Barvinok2016}.
Performing two recursive steps and excluding the hafnian of an auxiliary matrix, one finds the following second-order recursive formula
\begin{equation} \label{recursion}
\begin{split}
  \haf \tilde{C}(n) &= (2n-1) \eta' \, \haf \tilde{C}(n-1) 
                    \\
                    &\qquad -(n-1)^2 \big((\eta')^2-|\alpha'|^2\big) \, \haf \tilde{C}(n-2).  
\end{split} 
\end{equation}
It starts from the plain values of the hafnian $\haf \tilde{C}(1) = \eta'$ at $n=1$ and $\haf \tilde{C}(0) = 1$ at $n=0$ (as per convention in the hafnian master theorem).
Comparing it with the well-known recursive relation for the Legendre polynomials \cite{AbrStg} $P_n$,
$$ n P_n (x) = (2n-1) x \,P_{n-1}(x) - (n-1) P_{n-1}(x),$$
we at once reduce the hafnian to the Legendre polynomials: 
\begin{equation}
\haf \tilde{C}(n) = n! \big((\eta')^2 - |\alpha'|^2\big)^{n/2}\,
                        P_n\bigg(\eta'\Big/\sqrt{(\eta')^2-|\alpha'|^2}\bigg).
\end{equation}

Thus, the atom-number probabilities for sampling the occupation of the single-mode-squeezed standing plane wave are as follows
\begin{equation} \label{rho=Legendre}
\begin{split}
    \rho_n  &= \frac{\big(\, (\eta')^2-|\alpha'|^2\,\big)^{n/2}}{\sqrt{\det(\mathbbm{1}+G^{({\bf k})})}} 
                P_n \left( \frac{\eta'}{\sqrt{(\eta')^2-|\alpha'|^2}}\right) 
    \\
    &=\frac{1}{\sqrt{(\eta+1)^2-|\alpha|^2}} 
                \, \left[\frac{\eta^2-|\alpha|^2}{(\eta+1)^2-|\alpha|^2}\right]^{n/2}
    \\
    & \qquad \times P_n \left( \frac{\eta^2+\eta-|\alpha|^2}
                                    {\sqrt{(\eta^2-|\alpha|^2)\big((\eta+1)^2-|\alpha|^2\big)}} \right).
\end{split}
\end{equation}
The second equality is due to substitution $(\eta')^2-|\alpha'|^2 = \big(\eta^2 - |\alpha|^2\big)\big/ \big((\eta+1)^2 - |\alpha|^2\big)$.
Note that all Legendre polynomials $P_n$ have the same, independent on the atom number $n$, argument which is determined only by the normal and anomalous correlators. Instead, the atom number $n$ appears in the order of the Legendre polynomials.

\begin{figure*} 
\includegraphics{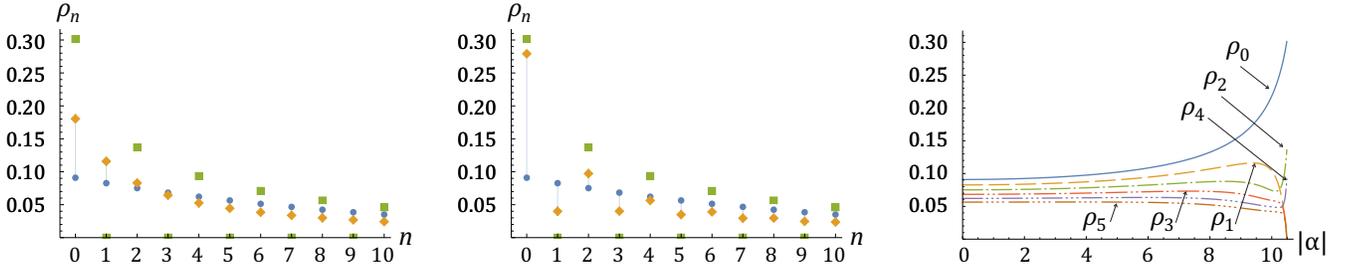}
\caption{   \label{fig:rho_esm}
Probability $\rho_n$ for sampling $n$ atoms in a single eigen-squeeze mode (\ref{sc}): Orange points correspond to the moderate values of the anomalous correlator $|\alpha| = 9.5$ (left panel) or $|\alpha| = 10.4$ (central panel); blue and green points on the left and central panels show probabilities $\rho_n$ in the case of zero ($\alpha = 0$) and maximal ($|\alpha | = \sqrt{\eta (1+\eta)} \simeq 10.5$) squeezing, respectively.  
The right panel exemplifies a dependence of the probabilities $\rho_0, \ldots, \rho_5$ on the anomalous correlator $|\alpha|$.
The normal correlator for all panels has the same value $\eta = 10$.
}
\end{figure*}

Alternatively, the hafnian in Eq.~(\ref{rho_n}) can be calculated in a straightforward way via elementary combinatorics as follows
\begin{equation} \label{haf=sum}
\haf \tilde{C}(n) = (\eta')^n \sum_{k=0}^{\lfloor n/2 \rfloor} \frac{(n!)^2}{4^k(n-2k)!(k!)^2} \left( \frac{\alpha'}{\eta'} \right)^{2k}.
\end{equation}
Here the symbol $\lfloor n/2 \rfloor$ stands for the largest integer less than or equal to $n/2$, and the essential argument is 
\begin{equation} \label{alpha'/eta'}
\begin{split}
    &|\alpha'|/\eta' = \big(e^{E_{\bf k}/T}-e^{-E_{\bf k}/T}\big)\sinh 2r_{\bf k}.    
\end{split}    
\end{equation}
According to the identity (3.137) in \cite{Gould}, the sum in Eq.~(\ref{haf=sum}) is proportional to the Legendre polynomial of the order $n$. Thus, the result stated above in Eq.~(\ref{rho=Legendre}) is rederived.

The recursive relation in Eq.~(\ref{recursion}) suggests the suppression of odd-occupation probabilities at small values of $\eta'<|\alpha'|$.
In the extreme case of zero mean occupation of quasiparticles, that is when $\eta'=0$, or $|\alpha| = \sqrt{\eta (1+\eta)}$, the recursive relation becomes purely two-step,
\begin{equation} \label{haf:eta=0}
    \haf \tilde{C}(n) = (n-1)^2 |\alpha'|^2 \, \haf \tilde{C}(n-2), \quad \eta' = 0.
\end{equation}
Eq.~(\ref{haf:eta=0}) yields the result 
\begin{equation}
\rho_{2m} = \frac{1}{\sqrt{\eta+1}}\frac{\big((2m-1)!!\big)^2}{(2m)!}\left(\frac{\eta}{\eta+1}\right)^m,
   \quad
   \rho_{2m+1}=0,
\end{equation}
which is equivalent (due to identities $\big((2m-1)!!\big)^2/(2m)! = (2m)!/4^m/(m!)^2$ and $\max |\alpha'| = \tanh r_{\bf k}$) to the one given by the reduction of the hafnian of a $(2\times 2)$-block-diagonal matrix to the square of the hafnian of just one of matrix's blocks,
\begin{equation}
\begin{split}
    \haf  \begin{bmatrix} \alpha'\, J_{n\times n} &                                                   \mathbb{0}_{n\times n}
                                \\
                                \mathbb{0}_{n\times n}       & 
                                                    \alpha'\, J_{n\times n} \\
                            \end{bmatrix}
    &=                        
    \Big(\haf \alpha'\, J_{n\times n} \Big)^2
    \\    
    &= \begin{cases}
			0, & \text{odd $n$}\\
            \left(\frac{n!}{2^{n/2} (n/2)!} (\alpha')^n \right)^2, & \text{even $n$} ,
		 \end{cases}   
\end{split}
\end{equation}
where the symbol $\mathbb{0}_{n_1\times n_2}$ denotes the $n_1\times n_2$ matrix with zero entries.

It leads precisely to the formula for the well-known occupation probabilities for a mode in the squeezed vacuum state \cite{Weedbrook2012, Barnett1996}, 
\begin{equation}
   \rho_{2m} = \frac{1}{\cosh r}\frac{(2m)!}{4^m (m!)^2} \tanh^{2m} r_{\bf k},
   \quad
   \rho_{2m+1}=0.
\end{equation}
Thus, all odd-occupation probabilities are zero, and only even occupation numbers show up in the atomic boson sampling. 
This is not surprising since the zero average quasiparticle occupation, $\tilde N_{\bf k} = 0$, means that the mode is in the vacuum Fock state $| 0 \rangle\langle 0|$ with respect to the quasiparticle operators. 

Typical occupation probability patterns given by Eq.~(\ref{rho=Legendre}) for sampling from eigen-squeeze bare-atom excited states (\ref{sc}) are illustrated in Fig.~\ref{fig:rho_esm}. 
In the range $|\alpha| \leq \alpha_{cr}$, where the absolute value of the anomalous correlator is less than the critical value $\alpha_{cr} = \eta$, an increasing absolute value of the anomalous correlator $\alpha$ leads to growing probabilities of occupations which are less than some number $n_0$ and suppressing probabilities of occupations which are larger than that number $n_0$.
Such a behavior drastically switches to a completely different behavior when the absolute value of the anomalous correlator exceeds the critical value, $|\alpha| > \alpha_{cr} = \eta$, at which the argument of the Legendre polynomial jumps from the pure real to pure imaginary values through the infinity. Namely, then the probabilities of even occupation numbers start to grow, while probabilities of odd occupation numbers start to drop down. 
At the maximal possible value of the anomalous correlator, $\max |\alpha| = \sqrt{\eta (1+\eta)}$, which corresponds to the squeezed vacuum state, $\tilde{N}_{\bf k} = 0$, all probabilities for even occupation numbers become zero.

The statistics of the eigen-squeeze-mode occupation can be deduced directly from its characteristic function $\Theta(z)$. It gives explicit information on the moments as well as ordinary, $\kappa_m$, and generating, $\tilde{\kappa}_m$ cumulants of the sampling probabilities via the Taylor series of its logarithm,
\begin{equation} \label{cumulants}
\ln \Theta (z) = \sum_{m=1}^\infty \tilde{\kappa}_m \frac{(z-1)^m}{m!} = \sum_{m=1}^\infty \kappa_m \frac{(\ln z)^m}{m!}. 
\end{equation}
This characteristic function is given by the general result in Eq.~(\ref{CFdet}) as follows
\begin{equation} \label{CF-sc}
\begin{split}
    \Theta (z) &= \frac{1}{\sqrt{\det \left( \mathbbm{1} + G^{({\bf k})} - z \, G^{({\bf k})}  \right)}} 
    \\
    = &\frac{1}{\sqrt{\det \left( \mathbbm{1} + G^{({\bf k})}  \right)}
        \sqrt{ \vphantom{\big|} 
            1 - z   \, \tr (PC) + z^2 \, \det (PC)
            }} \ .
\end{split}
\end{equation}
In Eq.~(\ref{CF-sc}) we provide two expressions for it. The first one is convenient for the cumulant analysis, while the second one -- for the straightforward calculation of probabilities. 
In particular, we see that the second expression (for its equivalent form, see \cite{Englert2002}) is proportional to the well-known generating function of the Legendre polynomials \cite{AbrStg} that gives another proof of Eq.~(\ref{rho=Legendre}).
The first expression yields the exact result for the generating cumulants, 
\begin{equation} \label{kappa_m}
\tilde\kappa_m = \frac{(m-1)!}{2} \tr \big(G^{({\bf k})}\big)^m = 
\frac{(m-1)!}{2} \big( \lambda_1^m + \lambda_2^m \big) ,
\end{equation}
where $\lambda_{1,2}$, Eq.~(\ref{lambda_12}), are the eigenvalues of the covariance matrix (\ref{Gk}). 
The value of the first generating cumulant $\tilde{\kappa}_{m=1}$ yields the mean occupation $\la n \ra \equiv \la \hat{s}_{\bf k}^\dagger \hat{s}_{\bf k} \ra = \eta$ whose thermal and quantum contributions are equal to $(e^{E_{\bf k}/T}-1)^{-1}\cosh 2r_{\bf k}$ and $\sinh^2 r_{\bf k}$, respectively.
Adding the value of the second generating cumulant, we immediately get the standard deviation, $\sigma^2 \equiv \tilde{\kappa}_{m=1} + \tilde{\kappa}_{m=2} = \eta^2 + \eta + |\alpha|$.
The higher moments of the atom-number probability distribution can be found in a similar way.

It is worth noting that the presence of the nontrivial anomalous correlator makes this characteristic function in Eq.~(\ref{CF-sc}) very different from the one, $\Theta_{IBG} (z) = [1 - (z -1) \eta]^{-1}$, associated with the atom-number sampling statistics in the limit of non-interacting, ideal Bogoliubov gas (IBG, $\alpha = 0$) when all excited atom modes remain in a non-squeezed state.

\section{Two-mode squeezing of the atomic boson sampling in the basis of traveling plane waves: Reduction of the hafnian to the permanent and hypergeometric function}

Let's see now how the sampling statistics is getting more complex due to adding the effect of interference on top of the effect of pure squeezing considered in the previous section. The minimal complication occurs when we choose the observational basis for sampling $\{ \phi_l \}$ (which is the set of bare-atom excited states for atom-number measurements) to be the traveling plane waves $\propto e^{\pm i{\bf kr}}$, Eq.~(\ref{expikr}). This is the simplest possible unitary mixing of the eigen-squeeze modes (\ref{sc}) employed as the observational basis in the previous section. So, the excited atoms are described now by the field operator (\ref{field-k}) via the annihilation operators $\hat{a}_{\bf k}, \hat{a}_{\bf -k}$ which differ from the annihilation operators $\hat{s}_{\bf k}, \hat{c}_{\bf k}$ of the eigen-squeeze modes due to the unitary transformation
\begin{equation} \label{V}
V = \frac{1}{\sqrt{2}}  \begin{bmatrix}     +i & -i \\
                                            +1 & +1
                                            \end{bmatrix},
    \qquad \begin{pmatrix} \hat{s}_{\bf k} \\  \hat{c}_{\bf k} \end{pmatrix} = V 
    \begin{pmatrix} \hat{a}_{\bf +k} \\  \hat{a}_{\bf -k} \end{pmatrix}.
\end{equation}
As a result, the Bogoliubov transformation to the eigen-energy quasiparticles in its Bloch-Messiah reduction form (\ref{BlochMessiah}) acquires, in addition to the central pure squeezing block $\tilde{R}_r$ in Eq.~(\ref{scBog}), the right-side unitary block $\tilde{R}_V$ with the unitary matrix $V$ which is given in Eq.~(\ref{V}) above and is not equal to the identity matrix anymore. The left-side unitary block in the Bloch-Messiah reduction (\ref{BlochMessiah}) remains trivial, $\tilde{R}_W = \mathbbm{1}$, since it is the unitary transformation from the eigen-squeeze two-component excitations to the eigen-energy quasiparticles (see Fig.~\ref{diag} in section IV) which in the case of the uniform BEC in a box trap coincide with the eigen-squeeze two-component excitations. The unitary transformation of the basis excited states corresponding to the operator transformation in Eq.~(\ref{V}) is performed by the unitary $V^*$ as per Eq.~(\ref{unitaryMixingV}), 
\begin{equation} \label{exp-sc}
    \sqrt{2}\begin{pmatrix}  \sin({\bf k r}) \\ \cos({\bf k r}) \end{pmatrix}
    =
     V^*
    \begin{pmatrix}  e^{+i\bf k r} \\ e^{-i\bf k r} \end{pmatrix}
    = 
    \frac{1}{\sqrt{2}}
    \begin{bmatrix}     -i & +i \\
                        +1 & +1         \end{bmatrix}
    \begin{pmatrix}  e^{+i\bf k r} \\ e^{-i\bf k r} \end{pmatrix}.
\end{equation}

Obviously, the introduced interference proceeds independently within each $(2\times 2)$-block of the atom excited states with the wave vectors ${\bf k}$ and $-{\bf k}$. Hence, it suffices to consider its effect on the sampling statistics just for one of such blocks --- the block corresponding to the following 4-vector of creation and annihilation operators $(\hat{a}_{\bf k}^\dagger, \hat{a}_{\bf -k}^\dagger, \hat{a}_{\bf k}, \hat{a}_{\bf -k})^\tp$. The related covariance matrix defined in Eq.~(\ref{G=x}) is equal to 
\begin{equation} \label{Gexp}
\begin{split}
&G^{({\bf k})}_{{\rm exp}} =\left[  \begin{matrix}    \eta    & 0         & 0         & \alpha   \\
                                0       & \eta      & \alpha  & 0         \\
                                0       & \alpha    & \eta      & 0         \\
                                \alpha  & 0         & 0         & \eta      \\
            \end{matrix} \right]
= -\frac{\mathbbm{1}}{2} + \left(\tilde{N}_{\bf k} + \frac{1}{2}\right) 
\\
& \ \ \times\left[  \begin{matrix}  \cosh 2r_{\bf k} & 0 & 0 & -\sinh 2r_{\bf k} \\
                            0 & \cosh 2r_{\bf k} & -\sinh 2r_{\bf k} & 0  \\
                            0 & -\sinh 2r_{\bf k} & \cosh 2r_{\bf k} & 0 \\
                            -\sinh 2r_{\bf k} & 0 & 0 & \cosh 2r_{\bf k} \\
            \end{matrix} \right] \! .
\end{split}
\end{equation}
It immediately follows from the general formula in Eq.~(\ref{GviaR}) after plugging in the Bloch-Messiah reduction of the Bogoliubov transformation stated above. 
Equivalently, it can be easily obtained by means of the unitary transformation $V$ from the covariance matrix in Eqs.~(\ref{GQ})-(\ref{CMcanon}) or (\ref{Gk}) written for the 4-vector $(\hat{c}_{\bf k}^\dagger,\hat{s}_{\bf k}^\dagger,\hat{c}_{\bf k},\hat{s}_{\bf k})^\tp$ of the creation and annihilation operators of the eigen-squeeze modes in the form of the $(4\times 4)$-matrix
\begin{equation}
\begin{split}   \label{G-sin-cos}
    &G^{({\bf k})}_{{\rm sin}} =\left[  \begin{matrix}    \eta    & 0         & \alpha  & 0         \\
                                0       & \eta      & 0         &\alpha    \\
                                \alpha  & 0         & \eta      &0   \\
                                0       & \alpha    & 0         &\eta
            \end{matrix} \right] 
    = -\frac{\mathbbm{1}}{2} + \left(\tilde{N}_{\bf k} + \frac{1}{2}\right) 
    \\ 
    & \ \ \times\left[  \begin{matrix}  \cosh 2r_{\bf k} & 0 & -\sinh 2r_{\bf k} & 0    \\
                                    0 & \cosh 2r_{\bf k} & 0 & -\sinh 2r_{\bf k}    \\
                                    -\sinh 2r_{\bf k} & 0 & \cosh 2r_{\bf k} & 0    \\
                                    0 & -\sinh 2r_{\bf k} & 0 & \cosh 2r_{\bf k}
            \end{matrix} \right] \! .
\end{split}
\end{equation}
The normal and anomalous correlators in the exponential-function basis remain the same as they were in the sinusoidal-function basis and are equal to $\eta$ and $\alpha$, respectively (see Eq.~(\ref{Gk})).

It is easy to find the covariance-related matrix in Eq.~(\ref{C}) via Eq.~(\ref{Gexp}) in the explicit form, 
\begin{equation}
    C \equiv P G^{({\bf k})}_{{\rm exp}} \Big(1+G^{({\bf k})}_{{\rm exp}}\Big)^{-1} = \begin{bmatrix}
            0 & \alpha' & \eta' & 0 \\
            \alpha' & 0 & 0 & \eta' \\
            \eta' & 0 & 0 & \alpha' \\
            0 & \eta' & \alpha' & 0
        \end{bmatrix},
\end{equation}
where the parameters $\eta', \alpha'$ are the entries of the covariance-related matrix $C$ in Eq.~(\ref{C-k}) describing the eigen-squeeze mode $\sqrt{2/V}\sin ({\bf kr})$ (see Eq.~(\ref{sc})). 

Now we can calculate the joint probability distribution of atom numbers in the two excited states which are the traveling plane waves $e^{i{\bf kr}}\big/\sqrt{V}$ and $e^{-i{\bf kr}}\big/\sqrt{V}$ for a given wave vector ${\bf k}$. 
The result is provided by the hafnian master theorem (\ref{pdf=Hafnian}),
\begin{equation} \label{rho_n1n2}
\begin{split}
    &\rho_{n_1,n_2} = \frac{\haf \tilde{C}(n_1,n_2)}{n_1!\,n_2!\, \big[(\eta+1)^2-|\alpha|^2\big]} \ ,
    \\
    &\tilde C(n_1,n_2) = \begin{bmatrix} 
            \mathbb{0}_{n_1\times n_1} & \alpha' J_{n_1\times n_2} 
            & \eta' J_{n_1\times n_1} & \mathbb{0}_{n_1\times n_2} \\
        \alpha' J_{n_2\times n_1} & \mathbb{0}_{n_2\times n_2} 
            & \mathbb{0}_{n_2\times n_1} & \eta' J_{n_2\times n_2} \\
        \eta' J_{n_1\times n_1} & \mathbb{0}_{n_1\times n_2} 
            & \mathbb{0}_{n_1\times n_1} & \alpha' J_{n_1\times n_2} \\
        \mathbb{0}_{n_2\times n_1} & \eta' J_{n_2\times n_2}
            & \alpha' J_{n_2\times n_1} & \mathbb{0}_{n_2\times n_2}
        \end{bmatrix}.
\end{split}        
\end{equation}
Here the hafnian matrix function is applied to the extended covariance-related matrix $\tilde C(n_1,n_2)$ which is a block matrix.  
The block $J_{n_1\times n_2}$ is a $n_1\times n_2$ matrix with all entries equal unity, $\mathbb{0}_{n_1\times n_2}$ is a zero matrix of the same size. The determinant in the denominator has been calculated explicitly as follows: $\det\big(\mathbbm{1}+G^{({\bf k})}_{{\rm exp}}\big) = \big[\,(\eta+1)^2-|\alpha|^2\,\big]^2$.

\begin{figure*} 
\includegraphics{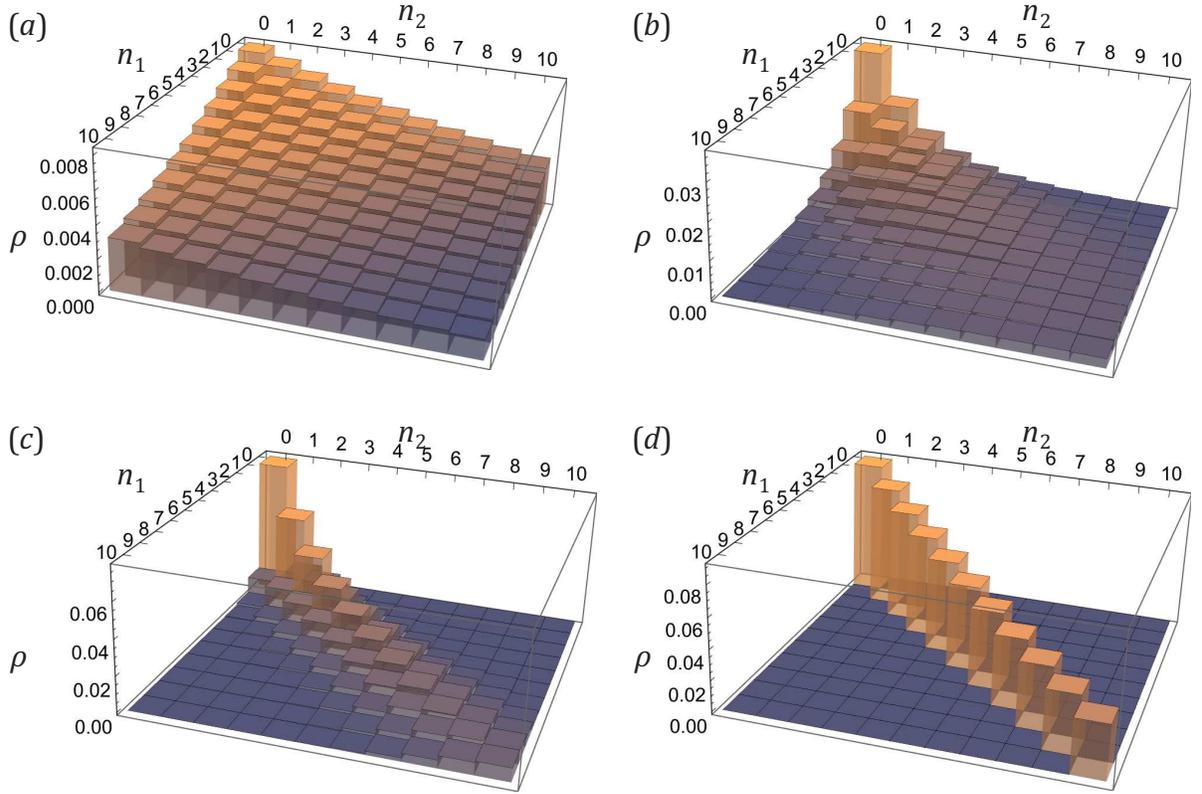}
\caption{   \label{fig:rho_2exp}
Typical dependence of the joint probability distribution $\rho_{n_1,n_2}$ for sampling $n_1$ and $n_2$ atoms in the two counter-propagating plane waves on the absolute value of the anomalous correlator, $|\alpha |$. The normal correlator $\eta$ (that is, the average number of atoms per one mode) is set to be $\eta$ = 10. The anomalous correlator values are (a) $\alpha = 0$ (which corresponds to an ideal, non-interacting gas), (b) $|\alpha | = 9.5$, (c) $|\alpha | = 10$, and (d) $\max|\alpha| = \sqrt{\eta (1+\eta)}$ (which corresponds to the two-mode squeezed vacuum state). 
Increasing $|\alpha|$ leads to growing up correlations between the random numbers $n_1$ and $n_2$. Note that the scale of the values $\rho_{n_1,n_2}$ changes significantly from (a) to (d).
}
\end{figure*}

The matrix $\tilde{C}(n_1,n_2)$ consists of blocks of unequal dimensions. However, its total dimension $2n_1+2n_2$ is even.
Such a hafnian $\haf \tilde{C}(n_1,n_2)$ is easy to compute.
The simplest way to do so is via consecutive swapping the $2$-nd and $4$-th block-rows and then the $2$-nd and $4$-th block-columns. 
This operation keeps the hafnian invariant, and allows us to convert the matrix into a block-counter-diagonal form for which the hafnian is reduced to the permanent as per the well-known identity
\begin{equation} \label{haf=per}
\haf \ \begin{bmatrix}  
            \mathbb{0} & A \\
            A^\tp & \mathbb{0}  \end{bmatrix}  
    = \textrm{per} \ A  
\end{equation}
valid for any square matrix $A$. As a result, we calculate the required hafnian as follows
\begin{equation} \label{hafC_n1n2}
\begin{split}
    &\haf \tilde C(n_1,n_2) =    \per \begin{bmatrix}  
            \eta' J_{n_1\times n_1} & \alpha' J_{n_1\times n_2} \\
            \alpha' J_{n_2\times n_1} & \eta' J_{n_2\times n_2}  \end{bmatrix} 
    \\
    &\quad=\sum_{k=0}^{\text{min}(n_1,n_2)} C_{n_1}^k C_{n_2}^k \, n_1! \, n_2! \,
                        (\eta')^{n_1+n_2-2k}|\alpha'|^{2k}
                    \\
    &\quad=\sum_{k=0}^{\text{min}(n_1,n_2)}
        \frac{(n_1!)^2 (n_2!)^2 (\eta')^{n_1+n_2-2k}|\alpha'|^{2k}}{(n_1-k)!\,(n_2-k)!\,(k!)^2} 
    \\    
    &\quad=n_1!\,n_2!\, (\eta')^{n_1+n_2} \ {}_{2}F_1\left(-n_1,-n_2,1,(\alpha'/\eta')^2\right).
\end{split} 
\end{equation}
Here the permanent has been calculated explicitly by simple combinatorial means via binomial coefficients $C_n^k = \frac{n!}{k!(n-k)!}$.
Indeed, one just calculates the number of permutations of $n_1+n_2$ elements which swap $k$ elements between the block of the first $n_1$ elements and the block of the last $n_2$ elements, which corresponds to the summand $\propto (\eta')^{n_1+n_2-2k}|\alpha'|^{2k}$.
The obtained sum is an ordinary (Gaussian) hypergeometric function ${}_{2}F_1$ with integer parameters which is a polynomial.
It could be also expressed in terms of the Jacobi polynomial \cite{Englert2002}, however, the hypergeometrical representation is more convenient. 
The argument of the hypergeometric function in terms of the eigen-energies $E_{\bf k}$ and eigen-squeezing parameters $r_{\bf k}$ is given in Eq.~(\ref{alpha'/eta'}).

The hafnian in Eq.~(\ref{rho_n1n2}) can be calculated also directly from its combinatorial definition \cite{Barvinok2016}. 
Each product of $n_1 + n_2$ entries contributing to the sum in the hafnian's definition is encoded by a division of the matrix dimension $2n_1+2n_2$ into $n_1+n_2$ unordered pairs.
Elements from the first group of $n_1$ elements could be paired either to the elements from the second group of $n_2$ elements, which corresponds to picking an $\alpha'$ entry from the non-diagonal block, or to the elements from the third group of $n_1$ elements, which corresponds to picking an $\eta'$ entry from the non-diagonal block.
Pairing to the same group or the fourth group means picking a zero element, which vanishes the summand.
Thus, each permutation encoding nonzero summand swaps $k$ elements between the first and second blocks (of $n_1$ and $n_2$ elements, respectively), as well as swaps $k$ elements between the third and fourth blocks. The rest of elements should be swapped between the first and third blocks, or between the second and fourth blocks.

Let $k \le \min(n_1,n_2)$ be a number of pairings between the first and second groups. 
It is also the number of pairings between the third and fourth groups.
There are $C_{n_1}^k C_{n_2}^k k!^2$ alternative ways to choose a pairing between the first and second groups, and each variant corresponds to the multiplier $(\alpha')^{k}$ accumulated from a non-diagonal block of $\tilde{C}$.
The same holds for pairing between the third and fourth groups.
Both the first and the third groups have $n_1-k$ unpaired elements in rest. There are $(n_1-k)!$ alternative ways to pair them correspondingly, each corresponds to the multiplier $(\eta')^{n_1-k}$.
Similarly, both the second and the fourth groups have $n_2-k$ unpaired elements in rest. There are $(n_2-k)!$ alternative ways to pair them correspondingly, each corresponds to the multiplier $(\eta')^{n_2-k}$.
Thus reaching the result stated above in Eq.~(\ref{hafC_n1n2}).

Combining Eqs.~(\ref{rho_n1n2}) and (\ref{hafC_n1n2}), we get the explicit formula for the joint probability distribution in the case of sampling from the atom excited states given by two counter-propagating traveling plane waves via the ordinary hypergeometric function:
\begin{equation} \label{rho_n1n2-hypergeom}
\begin{split}
    \rho_{n_1,n_2} &= \frac{n_1!\,n_2!\, (\eta')^{n_1+n_2}}{(\eta+1)^2-|\alpha|^2}
        \sum_{k=0}^{\text{min}(n_1,n_2)} \frac{(\alpha'/\eta')^{2k}}{(n_1-k)!\,(n_2-k)!\,(k!)^2}
    \\
    &\quad=\frac{(\eta')^{n_1+n_2}}{(\eta+1)^2-|\alpha|^2} \
        {}_{2}F_1\left(-n_1,-n_2,1,(\alpha'/\eta')^2\right).
\end{split}
\end{equation} 
An equivalent result may be derived via a straightforward differentiation of the characteristic function, but the corresponding calculations are more cumbersome \cite{Englert2002}. 

In the particular case of an ideal, non-interacting BEC gas, when the anomalous correlator equals zero, $\alpha = \alpha' = 0$, and the excited atom states are not squeezed, the calculation of the joint probability distribution in Eq.~(\ref{rho_n1n2}) becomes elementary since the hafnian of the corresponding, extremely degenerate matrix $\tilde C(n_1,n_2)$ is reduced to the product of two permanents: 
\begin{equation} 
\begin{split}
    &\rho_{n_1,n_2} = \frac{\per (\eta' J_{n_1\times n_1})
                \per (\eta' J_{n_2\times n_2})}{n_1!\,n_2!\,(\eta+1)^2} =
    \\
    &\qquad \frac{1}{\eta+1} \left(\frac{\eta}{\eta+1} \right)^{n_1}
                \times
                \frac{1}{\eta+1} \left(\frac{\eta}{\eta+1} \right)^{n_2}.
\end{split}        
\end{equation}
Thus, the joint probability distribution for occupations of the two degenerate counter-propagating waves separates into the product of two independent exponential single-mode distributions similar to the thermal counting statistics, Eq.~(\ref{rho_n=exp}).

In the presence of nonzero anomalous correlators the joint probability distribution could not be factorized anymore. This is due to appearance of the intra-modal squeezing. 
As is illustrated in Fig.~\ref{fig:rho_2exp}, increasing absolute value of the anomalous correlator results in enhanced correlations between the sampled atom numbers $n_1$ and $n_2$ and growing up probability of the completely entangled occupation states $n_1=n_2$. This occurs in accord with a simultaneous increase of the pure quantum contribution $G_Q$, Eq.~(\ref{GQ}), associated with the quantum depletion of the condensate and growing up single-mode squeezing parameter $r_{\bf k}$, to the covariance matrix in Eq.~(\ref{CMcanon}). The point is that the thermal quasiparticle occupation $\tilde{N}_{\bf k}$ and, hence, the relative value of the complimentary thermal contribution $G_T$, Eq.~(\ref{GT}), tend to zero when the anomalous correlator approaches it maximum value as per Fig.~\ref{fig:parameters}. As a result, the evolution of the probability distribution pattern in Fig.~\ref{fig:rho_2exp} looks like formation and appearance of the quantum entangled ridge out of the disappearing thermal covering layer, or background. 
Obviously, a full complexity in the sampling probability patterns becomes clearly visible when the relevant excited bare-atom ${\bf k}$-modes have a close-to-maximum anomalous correlator $|\alpha|$ and small mean quasiparticle occupation $\tilde{N}_{\bf k}$. Achieving this in the BEC experiments requires a proper choice of the sampling ${\bf k}$-modes as well as increasing the interatomic interaction, that is, the quantum depletion of the condensate, and decreasing the temperature, respectively.  

The leading term in the asymptotics of the joint probability distribution in Eq.~(\ref{rho_n1n2-hypergeom}) when the anomalous correlator approaches its extremum $\textrm{max} |\alpha| = \sqrt{\eta (1+\eta)}$ can be calculated directly from the hafnian formula in Eq.~(\ref{rho_n1n2}) by means of setting $\eta' = 0$ and employing the identity (\ref{haf=per}) as follows
\begin{equation} \label{rho_n1n2-asymp}
\begin{split}
    \rho_{n_1,n_2} & \sim \frac{1}{n_1!\,n_2!\,[(\eta+1)^2-|\alpha|^2]} 
        \per    \begin{bmatrix}  
                    \mathbb{0}_{n_1\times n_1} & \alpha' J_{n_1\times n_2} \\
                    \alpha' J_{n_2\times n_1} & \mathbb{0}_{n_2\times n_2}  \end{bmatrix}
    \\                    
    & = \frac{\delta_{n_1,n_2} (\alpha')^{n_1+n_2}}{(\eta+1)^2-|\alpha|^2}.
\end{split}
\end{equation}
It coincides with the contribution of the highest-order monomial in the hypergeometric-function polynomial in Eq.~(\ref{rho_n1n2-hypergeom}) and yields purely diagonal joint probability distribution in which the probability $\rho_{n_1,n_2=n_1}$ exponentially decreases with increasing atom numbers $n_1=n_2$. 

A well-known case of the squeezed vacuum state corresponds to the point at the very expremum, $|\alpha| = \textrm{max} |\alpha|, \eta' = 0$, when the asymptotics (\ref{rho_n1n2-asymp}) is reduced to a well-known result for the two-mode squeezed vacuum state \cite{Weedbrook2012, Barnett1996}
\begin{equation} \label{2modesqueezedvacuum} 
\rho_{n_1,n_2} = \delta_{n_1,n_2} \cosh^{-2}r_{\bf k}\, \tanh^{n_1+n_2} r_{\bf k} .
\end{equation}


\section{Nontrivial effect of interference on the atomic boson sampling in the basis of any two-mode-squeezed unitary-mixed degenerate standing plane waves}

Let's look at further complication of the sampling probability patterns due to nontrivial effect of interference in a more general set of the observational excited atom states. Consider atomic boson sampling from two excited atom states formed from the two eigen-squeeze modes $\sin({\bf kr})$, $\cos({\bf kr})$ by means of an arbitrary unitary mixing via the unitary matrix 
\begin{equation} \label{unitary2mode}
    V = 
        \begin{bmatrix}     \cos \xi    &   - e^{i \beta}\sin \xi \\
                            e^{-i \beta} \sin \xi   &   \cos \xi
            \end{bmatrix}, 
\qquad \begin{pmatrix} \hat{s}_{\bf k} \\   \hat{c}_{\bf k} \end{pmatrix} = V 
    \begin{pmatrix} \hat{b}_{{\bf k},1} \\  \hat{b}_{{\bf k},2} \end{pmatrix}. 
\end{equation}
It corresponds to the first part of the Bogoliubov transformation in Fig.~\ref{diag}, from the annihilation operators $\hat{b}_{{\bf k},1}, \hat{b}_{{\bf k},2}$ of these two excited atom states to the annihilation operators $\hat{s}_{\bf k}, \hat{c}_{\bf k}$ of the two eigen-squeeze modes.
Contrary to the particular case of two counter-propagating plane waves in Eq.~(\ref{V}), now the unitary $V$ involves two arbitrary real-valued angles $\xi$ and $\beta$. 
The matrix in Eq.~(\ref{unitary2mode}) has only two free parameters opposite to an arbitrary unitary matrix, 
$
    U = 
        \left[  \begin{smallmatrix}  x & y \\
                                -e^{i\gamma}y^* & e^{i\gamma}x^* \\
            \end{smallmatrix} \right],
$
$|x|^2+|y|^2 = 1$, $\gamma \in R$, involving four parameters.
Here the number of parameters is halved since the gauge phase factors of the two excited wave functions chosen to constitute the observational basis do not play any physical role.
In fact, the whole set of observational bases which correspond to physically different joint statistical distributions of atom numbers is parameterized by matrix $V$ in the form of Eq.~(\ref{unitary2mode}) with 
$\xi \in [0,\frac{\pi}{4}]$, $\beta \in [0,\frac{\pi}{2}]$.
In view of Eq.~(\ref{unitaryMixingV}), wave functions of the selected measurement basis,
\begin{equation}
    \begin{pmatrix}
        \phi_{{\bf k},1} ({\bf r}) \\ \phi_{{\bf k},2} ({\bf r})
    \end{pmatrix}    
    =
    V^\tp
    \begin{pmatrix}
        \sqrt{2/V}\sin ({\bf kr}) \\  \sqrt{2/V}\cos ({\bf kr})
    \end{pmatrix},
\end{equation}
are the superpositions of standing and traveling plane waves.

The covariance matrix for the operators $\hat{b}_{{\bf k},1}, \hat{b}_{{\bf k},2}$ could be easily obtained from the covariance matrix (\ref{G-sin-cos}) formed by the operators of the eigen-squeeze sine and cosine modes via the following transformation generated by the unitary $V$:
\begin{equation}
    G =     \left[  \begin{matrix}  V^\tp & \mathbb{0}   \\
                                    \mathbb{0} &  V^\dagger   
            \end{matrix} \right]
            \left[  \begin{matrix}  \eta \mathbbm{1}     &   \alpha \mathbbm{1} \\
                                    \alpha\mathbbm{1}   &    \eta\mathbbm{1}   
            \end{matrix} \right]
            \left[  \begin{matrix}  V^* & \mathbb{0}   \\
                                    \mathbb{0} &   V
            \end{matrix} \right]
        =   \left[  \begin{matrix}  \eta \mathbbm{1}        &   \alpha V^\tp V   \\
                                    \alpha V^\dagger V^*    &   \eta \mathbbm{1}  
            \end{matrix} \right].
\end{equation}
Here the diagonal block containing normal commutators is proportional to the identity matrix, while the unitary-induced interference affects the block with anomalous correlators.
The covariance-related matrix, which determines the matrix in the hafnian master theorem, takes the following form
\begin{equation} \label{C-unitary}
    C \equiv PG(\mathbbm{1}+G)^{-1} = 
        \begin{pmatrix}     \alpha' \, V^\dagger V^*    &   \eta' \mathbbm{1}   \\
                            \eta'   \mathbbm{1}         &   \alpha'\, V^\tp V
            \end{pmatrix} .
\end{equation}
Here the amplitudes $\alpha'$ and $\eta'$ (see Eq.~(\ref{C-k})) are the same as in sections V, VI.

Note that if the $2\times2$ symmetric matrix $V^\tp V$ appeared in Eq.~(\ref{C-unitary}) is proportional to the identity matrix and, hence, the same is true for the complex conjugated matrix $V^\dagger V^*$, computing probabilities is reduced to the simplest case of two independent eigen-squeeze modes solved in section V.
This case corresponds to selecting two orthogonal standing waves as a measurement basis.
In particular, such simplification happens if the matrix $V$ is chosen to be orthogonal, so that $V=V^*$ and $V^\tp V = \mathbbm{1}$.

The sampling probabilities are given by the hafnian master theorem (\ref{pdf=Hafnian}),
\begin{widetext}
\begin{equation}
\begin{split}       \label{rho2modes-deg-V}
    \rho_{n_1,n_2} = \frac{\haf \tilde{C}(n_1,n_2)}{n_1!\,n_2!\, \big((\eta+1)^2-|\alpha|^2\big)} \ ,
    \qquad
    \tilde C(n_1,n_2) = \begin{bmatrix} 
        \alpha'c_{1}^*\,J_{n_1\times n_1} & \alpha'c_{2}^*\,J_{n_1\times n_2} 
            & \eta' J_{n_1\times n_1} & \mathbb{0}_{n_1\times n_2} \\
        \alpha'c_{2}^*\,J_{n_2\times n_1} & \alpha'c_{3}^*\,J_{n_2\times n_2} 
            & \mathbb{0}_{n_2\times n_1} & \eta' J_{n_2\times n_2} \\
        \eta'\,J_{n_1\times n_1} & \mathbb{0}_{n_1\times n_2} 
            & \alpha'c_1\,J_{n_1\times n_1} & \alpha'c_2 J_{n_1\times n_2} \\
        \mathbb{0}_{n_2\times n_1} & \eta' J_{n_2\times n_2}
            & \alpha'c_2 J_{n_2\times n_1} & \alpha'c_3\,J_{n_2\times n_2}
        \end{bmatrix}\!.
\end{split}        
\end{equation}
\end{widetext}
In the present case of a general-type unitary mixing the hafnian in Eq.~(\ref{rho2modes-deg-V}) strongly depends on the variable entries of the nontrivial symmetric matrices $V^\tp V$ and $V^\dagger V^*$ in Eq.~(\ref{C-unitary}). Let's denote these entries as follows
\begin{equation}
\begin{split}
    &V^\tp V     = \begin{bmatrix}  c_{1}  &   c_{2}  \\
                                    c_{2}  &   c_{3}  \end{bmatrix},
    \qquad
    V^\dagger V^*   = \begin{bmatrix}   c_{1}^*  &   c_{2}^*  \\
                                        c_{2}^*  &   c_{3}^*  \end{bmatrix},
    \\     
    &c_{1} = \cos^2 \xi + e^{-2i\beta}\sin^2 \xi,
    \  c_{2} = -2i\cos\xi \sin\xi \sin \beta,
    \  c_{3} = c_{1}^*.
\end{split}    
\end{equation}

We can calculate the hafnian directly from its combinatorial definition implementing its further reduction to a combinatorial problem of counting different partitions of $2n_1+2n_2$ elements belonging to one of the four different groups (corresponding to different entries on the main diagonal) into pairs, either by linking an element from one group with an element from another group or by pairing elements within the same group.
Finally, we get the hafnian as the following sum
\begin{equation}    \label{haf2modes-deg-V}
\begin{split}
    &\haf \tilde{C}(n_1,n_2) =
    (n_1!n_2!)^2
    \sum_{d_1=0}^{n_1} \sum_{d_2=0}^{n_2} 
    \frac{(\eta')^{d_1+d_2} (\alpha')^{n_1+n_2-d_1-d_2}}{ d_1! \, d_2!} 
    \\
    &\ \times
    \sum_{j_1,j_3=0}^{\min(n_1-d_1,n_2-d_2)}
    \frac{(c_2)^{j_1}}{j_1!}
    \frac{(c_2^*)^{j_3}}{j_3!}
    \\
    &\ \times
    \frac{ \Big(\frac{c_1}{2}\Big)^\frac{n_1-d_1-j_1}{2} }{(\frac{n_1-d_1-j_1}{2})!}
    \frac{ \Big(\frac{c_3}{2}\Big)^\frac{n_2-d_2-j_1}{2} }{(\frac{n_2-d_2-j_1}{2})!}
    \frac{ \Big(\frac{c_1^*}{2}\Big)^\frac{n_1-d_1-j_3}{2} }{(\frac{n_1-d_1-j_3}{2})!}
    \frac{ \Big(\frac{c_3^*}{2}\Big)^\frac{n_2-d_2-j_3}{2} }{(\frac{n_2-d_2-j_3}{2})!} \ .
\end{split}
\end{equation}
It runs only over such subsets of indices which yield integer numbers under factorials in the denominator of Eq.~(\ref{haf2modes-deg-V}); otherwise, the summand should not be included in the sum.
In other words, all numbers $n_1-d_1-j_1$ and $n_2-d_2-j_1$ and $n_1-d_1-j_3$ and $n_2-d_2-j_3$ should be even.

\begin{figure*} 
\includegraphics{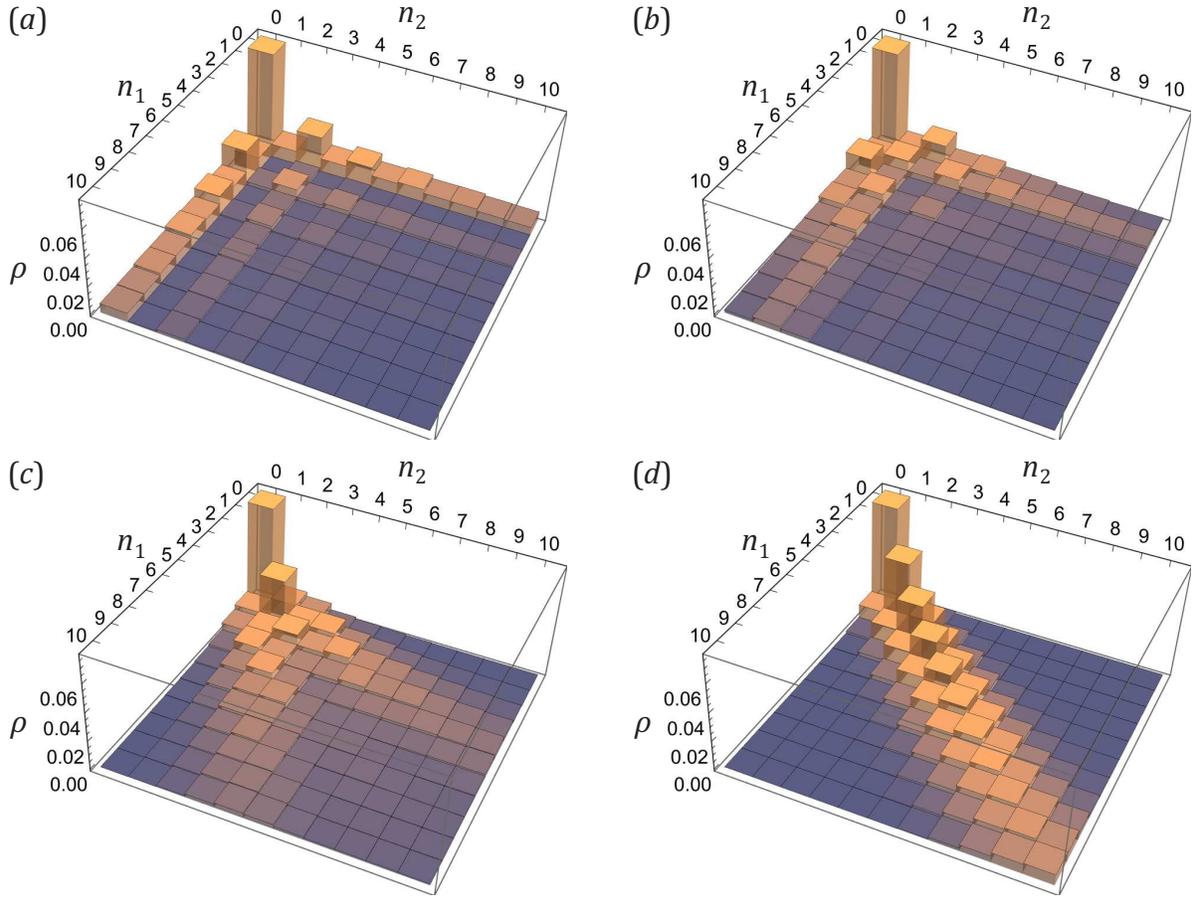}
\caption{   \label{fig:rho_unitary}
The joint probability distribution $\rho_{n_1,n_2}$ for atomic boson sampling from two excited atom states formed by a general-case unitary mixing of two eigen-squeeze modes (\ref{sc}) with the same wave vector ${\bf k}$.
The normal and anomalous correlators in Eq.~(\ref{Gk}) are $\eta =10$ and $|\alpha| = 10.4$, respectively, which amounts to the mean quasiparticle occupation $\tilde{N}_{\bf k} \simeq 0.95$ and the single-mode squeezing parameter $r_{\bf k} \simeq 1.34$.
The panels show evolution of the joint probability distribution with adjusting the matrix $V$ of the unitary mixing in Eq.~(\ref{BlochMessiah}):
(a) $V      =    \begin{bmatrix}    1   &   0   \\
                                    0   &   1   \end{bmatrix}$ 
(the sampling states coincide with the eigen-squeeze modes and the probability distribution factorizes into the product of two single mode distributions given in Eq.~(\ref{rho=Legendre})),
(b) $V  =   \dfrac{1}{2\sqrt{2}} \begin{bmatrix}   \sqrt{3}+1 & (1-\sqrt{3})i   \\
                                    (1-\sqrt{3})i & \sqrt{3}+1   \end{bmatrix}$,
(c) $V  =    \dfrac{1}{2}   \begin{bmatrix}     \sqrt{3} & -i   \\
                                                -i & \sqrt{3}   \end{bmatrix}$,
(d) $V  =    \dfrac{1}{\sqrt{2}} \begin{bmatrix}    +1 & -i   \\
                                                    -i & +1   \end{bmatrix}$ 
(the sampling states coincide with the two counter-propagating plane waves and the probability distribution is given by the hypergeometric function in Eq.~(\ref{rho_n1n2-hypergeom})).
The scaling on all panels is the same because the most probable outcome $n_1=n_2=0$ represents a Fock state invariant under unitary transformations.
}
\end{figure*}

Fig.~\ref{fig:rho_unitary} exemplifies how correlations between occupation numbers $n_1$ and $n_2$ arise while one varies the unitary $V$ making each observational-basis bare-atom excited state more inter-correlated combination of the eigen-squeeze modes. 
Occupation statistics of two eigen-squeeze modes is separable (see panel (a)). 
While one switches the observational basis from the eigen-squeeze standing waves to a basis consisting of partially-traveling waves, there appear nontrivial regions of the enlarged probabilities for some pairs of atom numbers $(n_1,n_2)$.
Switching to purely traveling waves, which is the case considered in the previous section, ultimately leads to forming a ridge extending along the diagonal $n_1=n_2$ direction (see panel (d)). 

In general, the probability pattern has a nontrivial structure determined by how the unitary mixing distributes the anomalous correlator, brought in by squeezing, over the entries $c_1, c_2$, and $c_3$. 
An example is a two-crest structure whose divergence angle depends on the unitary angles $\xi, \beta$ (see panels (b), (c)). 
It is strongly pronounced for the anomalous correlator values $|\alpha|$ close to its extremum, $\max |\alpha| = \sqrt{\eta (1+\eta)}$, since then the pure quantum, condensate-depletion-based contribution $G_Q$, Eq.~(\ref{GQ}), to the covariance matrix (\ref{CMcanon}) dominates the thermal one $G_T$, Eq.~(\ref{GT}).
Otherwise, for smaller $|\alpha|$ and larger $\eta$, the nontrivial correlation pattern is masked by thermal contributions due to large thermal population $\tilde{N}_{\bf k}$ of quasiparticles and small single-mode squeezing parameter $r_{\bf k}$ as per Fig.~\ref{fig:parameters}.

All of the probability distribution patterns on the plane of stochastic variables $n_1$ and $n_2$ shown in Fig.~\ref{fig:rho_unitary} are symmetric because we consider here the unitary mixture of two degenerate eigen-squeeze modes.
They have the same entries $\eta'$ and $|\alpha'|$ in their covariance-related matrices (see Eq.~(\ref{C-k})), which implies that the "diagonal" anomalous correlators have the same absolute value for any orthogonal states of the observational basis, in particular, $c_1=c_3^*$.

For a squeezed vacuum state, which implies zero mean quasiparticle occupation, the off-diagonal super-block of the extended covariance-related matrix $\tilde{C}$, Eq.~(\ref{rho2modes-deg-V}), is zero since $\eta'=0$.
Thus, the hafnian determining the probabilities is simplified to the product of the hafnians of two diagonal blocks which are complex conjugated to each other,
\begin{equation}
\begin{split}   \label{rho-2mixedmode-SV}
    &\haf \tilde{C}_{n_1,n_2}
    =
    \left|  \haf
        \begin{bmatrix}
            \alpha'c_1\,J_{n_1\times n_1} & 
                \alpha'c_2\,J_{n_1\times n_2} \\
            \alpha'c_2\,J_{n_2\times n_1} & 
                \alpha'c_3\,J_{n_2\times n_2} 
        \end{bmatrix}
    \right|^2
    \\
    & \ \ = (n_1!n_2!)^2 |\alpha'|^{n_1+n_2}
    \left|
        \sum_{k=0}^{\lfloor \mu/2 \rfloor}
        \frac{c_2^{\mu-2k} (c_1 c_3/4)^k}{k! (\mu-2k)! \left(k+|n_2-n_1|/2\right)!}
    \right|^2    
    \\
    & \ \ \times   \left|\frac{c_1}{2}\right|^{2\max(n_1-n_2,0)}
                    \left|\frac{c_3}{2}\right|^{2\max(n_2-n_1,0)},
    \quad \mu \equiv \min(n_1,n_2),
\end{split}
\end{equation}
for even $n_1+n_2$, and zero otherwise.
Here $\lfloor \mu/2 \rfloor$ stands for the maximal integer less or equal $\mu/2$, where $\mu$ is a minimal of two atom numbers $n_1$ and $n_2$.
Such a case of unitary mixed squeezed vacuum in optics had been described in \cite{Schrade1993} via integrals of Hermite polynomials, and the probability distribution had been finally reduced to the associated Legendre polynomial which parameters are determined by $n_1$ and $n_2$.
Note that in the squeezed vacuum state the probability of sampling the atom numbers $n_1$ and $n_2$ of different parity is always zero. This is an immediate consequence of the fact that the hafnian of the matrix of odd size $n_1+n_2$ is identically equal to zero. 

\begin{figure} 
\includegraphics{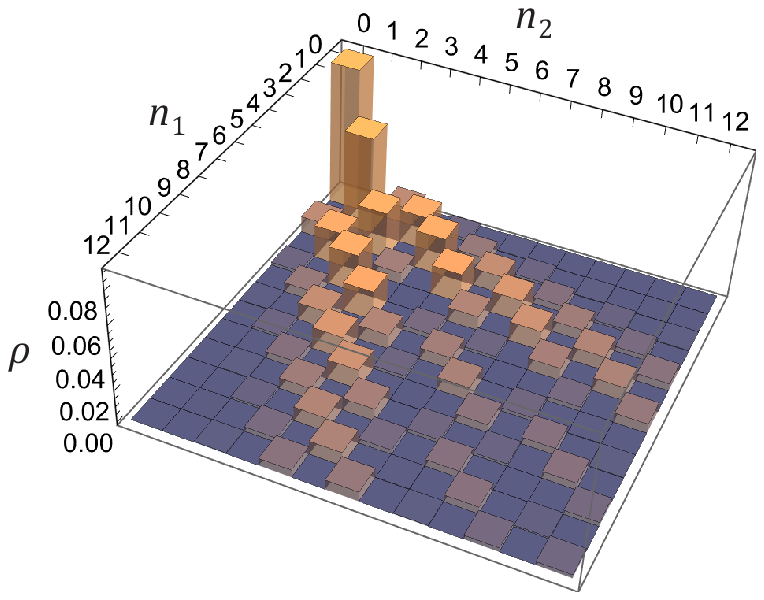}
\caption{   \label{fig:2MSV}
The joint probability distribution $\rho_{n_1,n_2}$ for atomic boson sampling from two excited atom states formed by a unitary mixing of two eigen-squeeze modes (\ref{sc}) with the same wave vector ${\bf k}$.
The unitary matrix is $V  =    \dfrac{1}{2}   \begin{bmatrix}     \sqrt{3} & -i   \\
                                                -i & \sqrt{3}   \end{bmatrix}$,
which coincides with the matrix corresponding to the panel (c) in Fig.~\ref{fig:rho_unitary}.
The system of excited atoms is in a squeezed vacuum state with zero quasiparticle occupations, $\tilde{N}_{\bf k} = 0$. The single-mode squeezing parameter, normal and anomalous correlators in Eq.~(\ref{Gk}) are $r_{\bf k} \simeq 1.87$, $\eta =10$ and $|\alpha| = \max |\alpha| = \sqrt{110}$, respectively.
}
\end{figure}

The squeezed vacuum state corresponds to the most contrasting correlation pattern of the joint probability distribution.
We illustrate this thesis by Fig.~\ref{fig:2MSV} plotted for the case of the vacuum state with the normal correlator $\eta=10$ and the unitary mixing matrix $V  =    \dfrac{1}{2}   \begin{bmatrix}     \sqrt{3} & -i   \\
                                                -i & \sqrt{3}   \end{bmatrix}$.
The wave functions of the selected measurement basis represent some mixtures of standing and traveling plane waves.
On the plane $(n_1,n_2)$ of sampling outcomes one can clearly see the dedicated directions along which the probabilities are relatively large.
This global landscape is also essentially decorated by a check-mate pattern representing the strong effect of probability cancellation for odd-parity total occupation stated above. 
The probabilities of adjacent outcomes typically dramatically differ from each other even if they are all close to the directions of large probabilities.  
There are even gaps of unlikely outcomes surrounded on all sides by outcomes with higher probabilities (for example, around the point $n_1=n_2=4$).

The presence of thermal excitations, i.e., the presence of a nonzero number of quasiparticles in the system, makes the correlation pattern not that sharp, as is seen from comparison of Fig.~\ref{fig:2MSV} and Fig.~\ref{fig:rho_unitary}, panel (c). 
Both of them refer to exactly the same matrix $V$ and observational basis, while the anomalous correlator $|\alpha|$ in Fig.~\ref{fig:2MSV} is larger.
The dedicated directions of more probable outcomes in Fig.~\ref{fig:rho_unitary} are still the same as in Fig.~\ref{fig:2MSV} and can be clearly seen, but they turn into gently sloping hills, and the abrupt checkerboard pattern has disappeared.

Note, however, that thermal excitations don't simply raise the background and the pattern of vacuum-state statistics doesn't just draw into them. 
One should take into account not just relative increase of the thermal contribution $G_T$, Eq.~(\ref{GT}), to the covariance matrix (\ref{CMcanon}) due to increasing quasiparticle population $\tilde{N}_{\bf k}$, but also decrease and restructuring of the pure quantum contribution $G_Q$, Eq.~(\ref{GQ}), due to simultaneous decrease of the single-mode squeezing parameter $r_{\bf k}$ as per Fig.~\ref{fig:parameters}, and mostly important a nontrivial mixing of the thermal and quantum contributions in the covariance matrix (\ref{CMcanon}) due to the unitary rotation $V$.  
The redistribution of probability which makes the landscape more smooth happens first among adjacent (or close enough) outcomes.
This is also seen from comparison of (d) panels in Fig.~\ref{fig:rho_2exp} and Fig.~\ref{fig:rho_unitary}, representing the joint occupation statistics for two counter-propagating plane waves for the states with no and few quasiparticles, respectively.
While in Fig.~\ref{fig:rho_2exp}(d), corresponding to the vacuum state, there is a well-pronounced and sharp diagonal ridge at $n_1=n_2$, the thermal excitations make it wider and enlarge, first of all, the probabilities of the adjacent states, $|n_1-n_2|=1,2$.
In general such behavior could be described by the formula for hafnians in Eq.~(\ref{haf2modes-deg-V}) which essentially allows one to employ a perturbation approach with respect to small value of the entries involving the parameter $\eta'$ (which are exactly zero for the vacuum state).

Importantly, while controlling the observational basis via the unitary $V$ may lead to nontrivial joint statistics of the atom numbers with widely ranging correlation patterns as described above, it doesn't affect the total noncondensate occupation statistics at all. This fact is not obvious from the formula for the joint probabilities since the direct sum representing the total-occupation probability,
\begin{equation}
    \rho_n \equiv \sum_{n_1=0}^n \rho_{n_1,n-n_1},
\end{equation}
may look nontrivial in virtue of Eq.~(\ref{rho2modes-deg-V}) involving complicated hafnians.

The distribution of the total number of atoms in any selected subset of excited states is easier to described via the characteristic function in Eq.~(\ref{CFdet}) with all arguments within this subset set to be equal, $z_j = z$. 
(Setting some groups of arguments equal to each other amounts to calculating a coarse-grained statistics.)
Then, the corresponding submatrix of variables $Z$ in Eq.~(\ref{CFdet}) turns into the scaled identity matrix  $z \mathbbm{1}$, which commutes with any matrix $V$. Thus, the unitary matrix, switching the observational basis and transforming the covariance matrix $G$, doesn't change the determinant in the expression for the characteristic function. 

In particular, for the considered case of mixing two eigen-squeeze modes the characteristic function of the total-occupation statistics is
\begin{equation}
    \theta(z,z) = \left[
                \det \left( \mathbbm{1} + 
                        \left[  \begin{matrix}  \eta    & \alpha  \\
                                                \alpha  & \eta  
                                \end{matrix} \right] \right) 
                \det \left( \mathbbm{1} - z 
                        \left[  \begin{matrix}  \eta' & \alpha'  \\
                                                \alpha' & \eta'  
                                \end{matrix} \right] \right)
                    \right]^{-1}.
\end{equation}
Its generating cumulants are easy to calculate as follows
\begin{equation}
\begin{split}
    &\tilde\kappa_m  = (m-1)! \tr G^m 
    \\
    &= (m-1)! 
                       \big( (\eta+|\alpha|)^m + (\eta-|\alpha|)^m \big)
    \\
    &=\Gamma(m) \left[
                    \left(\tilde N_{\bf k}e^{r_{\bf k}} + \frac{e^{r_{\bf k}}-1}{2}\right)^m
                    +
                    \left(\tilde N_{\bf k}e^{-r_{\bf k}} + \frac{e^{-r_{\bf k}}-1}{2}\right)^m
                \right],
\end{split}
\end{equation}
where $\Gamma$ denotes a gamma-function. 
They are exactly the same, up to an obvious common factor of 2, as the generating cumulants of a total occupation in independent eigen-squeeze modes with sinusoidal wave functions.
Knowing these cumulants, one may restore in a simple way the central moments of the distribution as per comments after Eq.~(\ref{kappa_m}) and detailed discussion in \cite{PRA2000}.

The probabilities of sampling $n$ atoms total in the considered excited states is easy to calculate expressing the determinant in the characteristic function in terms of the eigenvalues $\lambda'_{1,2}$ of the renormalized covariance matrix $G^{\bf(k)}\big(1+G^{\bf(k)}\big)^{-1}$,
\begin{equation}
\begin{split}
    &\theta(z,z) =  \frac{1}{    
                    \big(1-z \lambda'_1\big)\,\big(1-z \lambda'_2\big)\,
                    \det \left( \mathbbm{1} + 
                        \left[  \begin{matrix}  \eta    & \alpha  \\
                                                \alpha  & \eta  
                                \end{matrix} \right] \right)
                    }
    \\            
    &\quad =    \left(
                \frac{\lambda'_1}{1-z\lambda'_1} - \frac{\lambda'_2}{1-z\lambda'_2}
                \right)
                \frac{1}{    
                    \left( \lambda'_1 - \lambda'_2 \right)
                    \det \left( \mathbbm{1} + 
                        \left[  \begin{matrix}  \eta    & \alpha  \\
                                                \alpha  & \eta  
                                \end{matrix} \right] \right)
                                }.
\end{split}        
\end{equation}
Recalling  Eq.~(\ref{lambda'}), $\lambda'_{1,2}=\eta'\pm|\alpha'|$, and taking into account the relation 
$|\alpha'| \det \left( \mathbbm{1} + 
                        \left[  \begin{smallmatrix}  \eta    & \alpha  \\
                                                    \alpha  & \eta  
                                \end{smallmatrix} \right] \right) = |\alpha|$
following from Eq.~(\ref{C-k}), we finally get a simple formula for the probability of sampling $n$ atoms total:
\begin{equation} \label{rho2modeVtotal}
    \rho_n =  \frac{(\lambda'_1)^{n+1} - (\lambda'_2)^{n+1}}{2|\alpha|},
    \quad
    \lambda'_{1,2}  = \frac{1 \pm e^{E_{\bf k}/T}\tanh r_{\bf k}}{e^{E_{\bf k}/T} \pm \tanh r_{\bf k}}.
\end{equation} 
This result has been obtained for a pair of counter-propagating plane waves in \cite{PRA2000,Englert2002}.
Note that the combinations of parameters $Z(\pm A_{\bf k})$ or $Y_{\pm}$ introduced in \cite{PRA2000} or \cite{Englert2002} via some algebraic manipulations are, in fact, nothing else but inverse eigenvalues of the renormalized covariance matrix $G^{\bf(k)}\big(1+G^{\bf(k)}\big)^{-1}$.

In fact, the total noncondensate occupation statistics does not inherit the sophisticated, related to the $\sharp$P-hardness behavior of the joint probability distribution. 
However, it inherits the nontrivial property associated with the parity effect discussed above.
While the first eigenvalue in Eq.~(\ref{rho2modeVtotal}) is positive for any values of the normal and anomalous correlators, $0<\lambda'_1 \le 1$, the second eigenvalue $\lambda'_2$ becomes negative when the absolute value of the anomalous correlator gets larger than the normal correlator, $|\alpha| \geq \eta$. 
Then the probabilities of sampling an even total number of excited atoms are getting enhanced.
For a particular choice of the observational basis consisting of standing plane waves (see section~V), this property may be interpreted as a simple consequence of the fact that both independent occupation numbers $n_1$ and $n_2$ have strongly suppressed probabilities to be odd.
For the observational basis consisting of traveling plane waves the joint statistics has a strong correlation at $n_1=n_2$ (see section~VI), and the total-occupation probability $\rho_n = \sum_{n_1=0}^n \rho_{n_1,n-n_1}$ hits this diagonal only for even $n$.
In the vacuum state, when the anomalous correlator achieves its maximal possible absolute value, $|\alpha| = \sqrt{\eta^2+\eta}$, we have $\lambda'_1 = -\lambda'_2 = \tanh r_{\bf k}$, and the probabilities of odd total occupations completely vanish.

\section{The nature of the $\sharp$P-hard complexity: Squeezing and interference of atom sampling states via their interplay with eigen-squeeze modes and eigen-energy quasiparticles}

Let us briefly overview the aforestated analysis of the $\sharp$P-hard problem of atomic boson sampling from an interacting BEC gas trapped in a box. The analysis is done by means of a new approach based on the recently found hafnian master theorem \cite{PRA2022,LAA2022}. We intentionally choose a textbook quantum many-body model aiming to explain how to use the hafnian approach as a regular method for dealing with various $\sharp$P-hard problems. We infer that an equilibrium BEC gas in a box with periodic boundary conditions is one of the simplest models which has a potential for demonstrating quantum supremacy over classical computing. It allows us to greatly simplify and clarify the general formulas and reveal an explicit analytical description of the mechanism leading to the $\sharp$P-hardness of computing quantum properties of many-body systems. 

The general theory is formulated in sections II-IV. The simple explicit formulas and their numerical illustrations are given in sections V-VII which show increasing complexity of the joint probability distribution of the occupations of excited atom states with growing up complexity of unitary mixing used to form those excited atom states out of the eigen-squeeze modes (\ref{sc}) (compare patterns in Figs.~\ref{fig:rho_esm}-\ref{fig:2MSV}). Of course, while this unitary mixing occurs separately in different low-dimensional blocks of eigen-squeeze modes the joint probability distribution remains easily computable by analytical, recursive or numerical means. Only with increasing dimensionality of those mixing blocks of eigen-squeeze modes computing sampling probabilities requires exponential time and becomes $\sharp$P-hard.    

We find that there are two ingredients of the $\sharp$P-hardness of the atomic boson sampling --- the squeezing and the interference. Moreover, there are two corresponding unique entities existing in the BEC gas of interacting atoms, the eigen-squeeze modes and the eigen-energy quasiparticles, which are directly responsible for the above-mentioned squeezing and interference (see the schematic diagram in Fig.~\ref{diag}, section IV). The eigen-squeeze modes are the eigenvectors of the Hermitian factor of the multimode squeeze matrix. They are the elementary, intrinsic carriers of its eigenvalues --- the single-mode squeezing parameters, Eq.~(\ref{W-eigen-squeezing}). The quasiparticles are the eigenvectors of the Hamiltonian, Eq.~(\ref{QPH}), and are described by two-component wave functions, Eq.~(\ref{field:ba-qp}). They are the elementary collective excitations carrying quanta of collective energy --- the eigenvalues of the Hamiltonian. 

As such each eigen-energy quasiparticle lives completely independent on other quasiparticles, but it is a superposition of many two-component eigen-squeezed quasiparticles formed by the irreducible, eigen-squeezing Bogoliubov transformation $\tilde{R}_r$, Eq.~(\ref{BlochMessiah}), from the one-component eigen-squeeze modes, each of which is itself a superposition of many excited bare-atom wave functions (see the schematic diagram in Fig.~\ref{diag}). As a result, the excited bare-atom wave functions are persistently interfere with each other, and the joint probability distribution of their atom numbers turns out to be $\sharp$P-hard for computing. 

Only in some very special cases the above-mentioned probability distribution can be computed faster than in exponential time. 
An example is the case when each bare-atom excited state chosen for atom number measurements by a multi-detector imaging system coincides with an eigen-squeeze mode which, at the same time, gives the same spatial profile for both components of the quasiparticle wave function.
Then the joint probability distribution turns into a separable product of the occupation probabilities of the single eigen-squeeze modes. 
Each of those probabilities describes a nontrivial squeezed-state statistics, but is computable via Legendre polynomials as per Eq.~(\ref{rho=Legendre}).

Another, more involved example is the case when the bare-atom excited states are chosen to be traveling plane waves. Then atomic boson sampling splits into independent samplings within separable blocks of two counter-propagating plane waves with the wave vectors ${\bf k}$ and ${\bf -k}$. The sampling probability distribution for each of such blocks shows statistics of two-mode squeezing, which is more complex, but again is easily computable via the ordinary hypergeometric function as per Eq.~(\ref{rho_n1n2-hypergeom}).
Note that the presence of squeezing makes the sampling statistics of Bose-atom numbers, even in the two-mode case, very different from and much more involved than the joint occupation statistics of purely interfering non-squeezed bosons, including a well-known two-mode Hong-Ou-Mandel statistics of just interfering bosons \cite{Kaufman2018,Aspect2015}.

One more, extreme example is the case when the squeezing is absent, say, due to the absence of interatomic interactions, like in an ideal Bose gas, or due to the absence of the condensate, like in a classical gas above the critical temperature.
In this case all single-squeezing parameters in Eq.~(\ref{Lambda_r}) vanish. Hence, the hafnian in Eq.~(\ref{pdf=Hafnian}) determining the joint sampling probabilities reduces to the permanent of a positive matrix since the extended covariance-related matrix has vanishing anomalous-correlator blocks and one can employ the Stockmeyer’s approximating algorithm\cite{Aaronson2013,LundPRL2014,Lund2015,Stockmeyer} for such a permanent.

A pivotal key for understanding the origin of the $\sharp$P-hardness of atomic boson sampling is provided by the irreducible Bloch-Messiah reduction of the Bogoliubov transformation into the product of three blocks, $\tilde{R}_W \tilde{R}_r \tilde{R}_V$, as per Eq.~(\ref{BlochMessiah}) and Fig.~\ref{diag}. 
Via its direct relation to the hafnian of the extended covariance-related matrix constituting the general result for the joint probability distribution in Eq.~(\ref{pdf=Hafnian}), the Bloch-Messiah reduction explicitly reveals the mechanism of squeezing and two mechanisms of interference responsible for the $\sharp$P-hard complexity. The squeezing is attributed to the single-mode squeezing block $\tilde{R}_r$ of the Bogoliubov transformation.
The eigenvalues $\{ r_l \}$ of the Hermitian factor of the multimode squeeze matrix, that is, the squeezing parameters of the eigen-squeeze modes, form an irreducible resource of the system of many interacting atoms in the BEC trap. 
The existence of squeezing in the BEC gas is known since \cite{PRA2000}. 
The interference, controlled by the block $\tilde{R}_V$ and its unitary $V$ as per Eq.~(\ref{unitaryMixingV}), between the observational bare-atom excited states and the eigen-squeeze modes constitutes the first mechanism of interference. 
The interference between the eigen-squeeze two-component excitations and the eigen-energy quasiparticles, controlled by the block $\tilde{R}_W$ and its unitary $W$ as per Eq.~(\ref{eigen-squeeze}), constitutes the second mechanism of interference (see the schematic diagram in Fig.~\ref{diag}). 

In other words, the interacting BEC gas in a trap has two naturally built-in, intrinsic interferometers associated with the two interference mechanisms disclosed above. 

On this basis, we conclude that even if just one of the interference mechanisms is available for controlling parameters of sampling in a wide range, then the $\sharp$P-hard complexity for the average case still exists. This is the case for the presented model of atomic boson sampling in a box trap with a uniform condensate for which the parameters of the quasiparticles and eigen-squeeze modes, including the unitary $W$ defining their interference block $\tilde{R}_W$, are almost fixed by a given trapping potential and couplings in Eq.~(\ref{HH}) and cannot be varied in a wide range. So, a wide variability of the BEC parameters and Bogoliubov couplings provided by the multi-qubit BEC trap \cite{Entropy2022} is useful, but not necessary for demonstrating computational $\sharp$P-hardness and potential quantum supremacy of atomic boson sampling. 

Mathematically, this $\sharp$P-hardness is the property of the hafnian (or permanent) \cite{Valiant1979,Jerrum2004,Bjorklund2019} of the extended covariance-related matrix in Eq.~(\ref{pdf=Hafnian}) which, according to the hafnian master theorem \cite{PRA2022,LAA2022}, determines the atom-number sampling probabilities. These probabilities are calculated as the Fourier series coefficients (\ref{PDF}) of the easy-to-compute characteristic function in Eq.~(\ref{CFdet}). Thus, the $\sharp$P-hardness is due to an intuitively obvious complexity of computing the multivariate Fourier integral in Eq.~(\ref{PDF}) for a sign-indefinite strongly-oscillating function (its analog is a lacunary or fractal function with an exponentially wide spectrum) \cite{PRA2022,Entropy2020}. 

The $\sharp$P-hardness of computing the hafnian in Eq.~(\ref{pdf=Hafnian})  follows \cite{Aaronson2013,HamiltonPRA2019} from two known facts: (a) the Haar randomness of the unitary matrices yields the Gaussian randomness of the extended covariance-related matrix $\tilde{C}(\{n_l\})$ and (b) computing the
hafnian of a random Gaussian matrix is a $\sharp$P-complete problem. 
The $\sharp$P-complete problems constitute the top-level complexity class within the class of $\sharp$P-hard problems. In virtue of the Toda's theorem \cite{Toda1991,Basu2012}, solution of any $\sharp$P-complete problem is reducible in polynomial time to the solution of any other problem in this class. Therefore, the multivariate Fourier integration can be viewed as the universal origin, or source, of the computational $\sharp$P-hardness and potential quantum supremacy of the many-body quantum systems. The point is that the quantum many-body systems process the multivariate Fourier-series transform naturally, as a routine part of their life, that is in linear or, at least, polynomial time, while the classical simulators or computers can do this only in exponential time. 

In fact, the multivariate Fourier transform described above reveals a certain duality of quantum and classical computational complexity. Both the characteristic function and its Fourier transform, that is its Fourier-series coefficients which constitute the joint probability distribution of sampling probabilities, contain full information about the atomic sampling statistics. However, the former, as any matrix-determinant function, can be easily calculated by classical computers while the latter cannot. There is no contradiction hidden in this statement since calculation of the multivariate Fourier integral for the sampling probabilities requires, in a general average case, computing the characteristic function under the integral in the exponentially large number of points. First, it means that there is nothing mysterious in $\sharp$P-hardness since it is just an ordinary property of multivariate integration. Second, it means that deriving the sampling probabilities from experimentally accumulated probabilities of atom-number samples, each of which is easily given by the BEC-gas quantum simulator almost in no time, requires an exponentially large number of samples (experimental runs) and, hence, an exponential time. Thus, only some special problems, such as a generation of strings of random numbers obeying the hafnian-based probability distribution (\ref{pdf=Hafnian}), are easy for the quantum many-body BEC-gas system but exponentially hard for classical computing.   

\section{Towards experiments on manifestations of $\sharp$P-hard complexity and quantum supremacy of atomic boson sampling}

We emphasize that, contrary to a widely discussed Gaussian boson sampling of noninteracting photons in a linear interferometer, the proposed atomic boson sampling does not require sophisticated synchronized external sources of bosons in squeezed states. The squeezing and interference of atom excited states, both of which are necessary for the computational $\sharp$P-hardness of boson sampling, are self-generated even in an equilibrium BEC gas. Hence, the major limitation factor for achieving quantum supremacy via boson sampling in a deep linear interferometer, which is an exponential loss of photons due to scattering and absorption on coupling elements (beam splitters, phase shifters, etc.) during propagation through the interferometer, is not an issue for the atomic boson sampling.  

Conceptually, the experiments on sampling are simple. In fact, the experiments on the statistics of the total occupation of excited atom states, that is the total noncondensate occupation, has been successfully performed \cite{Rzazewski2019}. In order to pioneer atomic boson sampling one just needs to split the noncondensate into fractions and to measure many times the atom occupation numbers in a preselected subset of excited wave functions by means of some multi-detector imaging system. Then, (a) to reconfigure detectors and/or trapping potential and other parameters of the BEC-gas in a trap and (b) to measure sampling statistics for such a unitary-transformed subset of excited wave functions and new system's parameters, and so on.  There is no need neither in any controllable non-equilibrium unitary-evolution processes typical for most quantum-computing experiments nor in suppression of various concomitant processes of relaxation and decoherence. The quantum system of interacting atoms in a BEC trap just simulates its own equilibrium life which consists of persistent quantum-statistical fluctuations. It is described by the statistical operator that intrinsically involves properties which are $\sharp$P-hard for computing. 

The $\sharp$P-hardness of computing the hafnian-based sampling statistics is a fundamental reason for, but of course not equivalent to or sufficient for quantum supremacy of the BEC gas in a trap over classical simulators with respect to generation the strings of random numbers obeying the predetermined joint probability distribution in Eq.~(\ref{pdf=Hafnian}). Whether a classical computer/simulator can or cannot provide generation of such random numbers in polynomial time is an open question. Obviously, the quantum system of many interacting atoms in a BEC trap is in a privileged position because its equilibrium state naturally provides the required sampling statistics. As is always the case in discussion of quantum supremacy, the very choice of the problem for simulation is intentionally unfair with regard to classical computing. Surely, the relation between classical and quantum computing is asymmetric.

The aforestated analysis of atomic boson sampling for the interacting atoms in the BEC trap, in particular, the result for sampling probabilities in Eq.~(\ref{pdf=Hafnian}), reveals its close similarity to the Gaussian boson sampling of noninteracting photons in a linear interferometer. 
This fact allows one to transfer an existing extensive analysis of the prospects and requirements for demonstrating quantum supremacy of photonic boson sampling 
\cite{Harrow2017, Zhong2020, Hamilton2017, HamiltonPRA2019, Lund2015, Quesada2022, LundPRL2014, Bentivegna2015, Shi2021, Chin2018, Quesada2018, Zhong2019, Brod2019, Huh2019, Huh2020, Wang2019, PanPRL2021, Madsen2022, Villalonga2021, BentivegnaBayesianTest2015, Renema2018, Renema2020, Popova2021, Qi2020} 
to the case of atomic boson sampling. 
For instance, two previously suggested schemes of photonic Gaussian boson sampling -- the scheme that smuggles a random Gaussian matrix as a submatrix of the covariance matrix \cite{Hamilton2017} and another scheme that smuggles an arbitrary symmetric matrix \cite{HamiltonPRA2019} -- could be, in principle, imitated within the atomic BEC platform by a proper choice of the system parameters and unitary $V$.
Demonstration of the recently suggested bipartite protocol of photonic Gaussian boson sampling \cite{Quesada2022} is also possible since the required balanced two-mode squeezing is naturally generated in the atomic BEC box trap for the pairs of degenerate counter-propagating plane waves $e^{\pm i {\bf kr}}$, see Eq.~(\ref{unitaryMixingV}).
Subsequent mixing of the first wave components of these pairs ($e^{+i {\bf kr}}$ waves) via a unitary $V_+$ and independent mixing of the second wave components of these pairs ($e^{-i {\bf kr}}$ waves) via a different unitary $V_-$ yield the bipartite protocol by involving a large-size, $\frac{M}{2}\times \frac{M}{2}$ matrix block $V_+ (\tanh \Lambda_r) V_-$ which could be set arbitrary in view of the theorem on the singular value decomposition of a matrix.
However, such an analysis goes beyond the scope of the present paper. 

We just note a recent analysis \cite{Aaronson2013,Hamilton2017,HamiltonPRA2019,Jiang2006,Jiang2009,Quesada2022,Lim2022} suggesting a possibility of achieving quantum supremacy in the regimes with the mean number of bosons per each of $M$ sampled modes (channels) on the order of $1/M^{1/2}$ (dilute sampling regime) or even $\sim 1$ (high-collision regime), rather than only in a deeply unitary-hiding regime with a very small mean occupation per mode, $\sim 1/M^{4/5}$, as had been assumed previously \cite{Aaronson2011,Harrow2017,Boixo2018}. 
Two well-known recent experiments on photonic boson sampling \cite{PanPRL2021,Madsen2022} had been also implemented at the order-of-unity mean occupation per mode, featuring up to 113 or 219 photon detection events out of a 144 or 216-mode photonic circuit, respectively.   
Hopefully, the atomic boson sampling and other experiments based on such a BEC platform for studying computational $\sharp$P-hardness in quantum many-body systems will become available soon. 
Especially promising in this regard could be experiments similar to the measurement of the full counting statistics of excited, noncondensed atoms in the momentum space after their release from a trap and subsequent free-fall expansion \cite{Clement2021,Clement2023}. 
Remarkably, in these experiments the detectors with a large quantum efficiency and single-atom resolution have been demonstrated. 

As is discussed above and illustrated in Figs.~\ref{fig:rho_2exp}--\ref{fig:2MSV}, a clear demonstration of a full complexity of the sampling probability patterns in the experiments with atomic BEC-gas requires a proper choice of the observational excited bare-atom states, a strong enough interatomic interaction (i.e., quantum depletion of the condensate) and low enough temperature so that the squeezing would be strongly pronounced and quantum statistics would not be hidden under thermal fluctuations.

Discussion of various closely related techniques for imaging of the local atom-number fluctuations and achieving nearly single-atom resolution in BEC-gas experiments can be found in \cite{Pit2011,Castin,Kaufman2018,Jacqmin2010,Armijo2010,RaizenBECstatisticsPRL2005,AspectDensityFluctPRL2006,Dotsenko2005,Schlosser2002,Raizen2009}. 

An ultimate demonstration of quantum supremacy for the average case \cite{Aaronson2013,Harrow2017,Boixo2018} can be achieved only if one gets an access to a wide-range unitary mixing, interference of the eigen-squeeze modes via either the observational basis states (the unitary $V$ in Eq.~(\ref{BlochMessiah})) or the quasiparticle states (the unitary $W$ in Eq.~(\ref{W-eigen-squeezing})). The former can be done solely by reconfiguring atom-number detectors for projecting onto the unitary mixed excited bare-atom states. Reconfiguration of the trapping potential and other parameters controlling the condensate profile and interatomic interactions provides control on the interference via both unitaries $W$ and $V$. In other words, one needs to get a control on either the first or second intrinsic interferometers built by nature in the interacting BEC gas which correspond to the first or second interference mechanisms revealed in sections~IV (see Fig.~\ref{diag}) and VIII. However, such a full control is not necessary for pioneering experiments on atomic boson sampling. Obtaining nontrivial patterns (see Fig.~\ref{fig:rho_2exp}) of the joint probability distribution of atom numbers for two counter-propagating traveling plane waves, for example, based on the full counting statistics accumulated in the available experiments \cite{Clement2021,Clement2023}, would be already a proof-of-principle demonstration of atomic boson sampling.   

Compared to the experiments on fluctuations of the total noncondensate occupation \cite{PRA2000,Rzazewski2019,PRA2020}, the experiments on atomic boson sampling do not imply counting all noncondensed atoms. In this respect the latter experiments are even simpler that the former ones since an exact separation of relatively small fraction of noncondensed atoms from much more occupied condensate, that is, drawing a precise borderline between the two, is the main challenge for the former experiments. For implementing atomic boson sampling, it suffices to measure joint occupation statistics just for some excited atom states (or coarse-grained groups of them) all of which could be far from the condensate wave function in the momentum space or, more generally, in the functional space and, therefore, easily distinguishable from the condensate. 

\begin{acknowledgments}
S. Tarasov acknowledges the support from the Foundation for the Advancement of Theoretical Physics and Mathematics “BASIS” (grant \#20-1-3-50-1).
\end{acknowledgments}

\subsection*{Data Availability} 
The data that support the findings of this study are available from the corresponding author upon reasonable request.

\subsection*{Conflict of interest} 
The authors have no conflicts to disclose. 

\nocite{*}

{}
\end{document}